 \def\versionno{ wall -- version 1.0 }
\catcode`\@=11

\expandafter\ifx\csname draftcontrol\endcsname\relax\global\def\draftcontrol{0}
\fi

{\count255=\time\divide\count255 by 60
\xdef\hourmin{\number\count255}
\multiply\count255 by-60\advance\count255 by\time
\xdef\hourmin{\hourmin:\ifnum\count255<10 0\fi\the\count255}}
\def\draftdate{\number\month/\number\day/\number\year\ \ \ \hourmin }

\newcommand\makepapertitle{\par
  \begingroup
    \renewcommand\thefootnote{\@fnsymbol\c@footnote}%
    \def\@makefnmark{\rlap{\@textsuperscript{\normalfont\@thefnmark}}}%
    \long\def\@makefntext##1{\parindent 1em\noindent
            \hb@xt@1.8em{%
                \hss\@textsuperscript{\normalfont\@thefnmark}}##1}%
     \newpage
     \global\@topnum\z@   % Prevents figures from going at top of page.
     \@makepapertitle
     \thispagestyle{empty}\@thanks
  \endgroup
  \setcounter{footnote}{0}%
  \global\let\thanks\relax
  \global\let\makepapertitle\relax
  \global\let\@makepapertitle\relax
  \global\let\@thanks\@empty
  \global\let\@author\@empty
  \global\let\@date\@empty
  \global\let\@title\@empty
  \global\let\title\relax
  \global\let\author\relax
  \global\let\date\relax
  \global\let\and\relax
  \def\version{\let\version\@version\@gobble}
}
\def\@makepapertitle{%
  \newpage
   \ifnum\draftcontrol=1 {}
   \version\versionno
   \vskip 3em%
   \else
   \hfill\hbox to 3cm {\parbox{4cm}{\@pubnum}\hss}%
   \vskip 3em%
   \fi
   \begin{center}%
   \let \footnote \thanks
     {\LARGE \@title \par}%
     \vskip 1.5em%
     {\normalsize%\large
       \lineskip .5em%
       \begin{center} %\begin{tabular}[t]{c}%
         \@author
       \end{center} % \end{tabular}
\par}%
     \vskip 1em%
     {\@bstract}%
     \end{center}%
     \vskip .5em
     \@date%
   \par
}

\gdef\@pubnum{}
\def\pubnum#1{%
  \gdef\@pubnum{#1}}

\gdef\@bstract{}
\def\Abstract#1{%
  \gdef\@bstract{%
   \parbox{\textwidth-0pc}{%
   \centerline{\bf Abstract}\penalty1000
   \noindent%\abstractfont \baselineskip=12pt
   \renewcommand\baselinestretch{1.0}
   {#1}}}
}

\def\ps@paper{\let\@mkboth\@gobbletwo%
     \ifnum\draftcontrol=1
        \def\@oddfoot{\hbox to \textwidth{\tiny \versionno \hfil\tiny\draftdate}%
        \hskip -\textwidth \hbox to \textwidth{\hfil\rm\thepage\hfil}}%
     \else\def\@oddfoot{\hbox to \textwidth{\hfil\rm\thepage\hfil}}
     \fi
     \let\@evenfoot\@oddfoot
}
\def\body{\clearpage
          \pagestyle{paper}
        }
\newenvironment{acknowledgments}{%
\vskip 3.25ex
\noindent {\bf Acknowledgments}
}

\def\@version#1{\ifnum\draftcontrol=1
\typeout{}\typeout{#1}\typeout{}
\vskip3mm\centerline{\hbox{\fbox{\normalsize{\tt DRAFT -- #1 -- }
                   {\draftdate}}}}\vskip3mm
\fi}
\let\version\@version
\long\def\eqlabel#1{\ifnum\draftcontrol=1
                    \tag@false  % there are some problems with multline without this
                    \tag*{(\theequation) \hbox to -0.2cm{\hspace{0cm}\small{#1}\hss}}
                    \refstepcounter{equation} 
                    \edef\@currentlabel{\theequation}
                    \ltx@label{#1}          % use old LaTeX \label instead of new definition
                    \else
                    \label{#1}
                    \fi
                    }
\let\st@bibitem\@bibitem
\let\st@lbibitem\@lbibitem
\ifnum\draftcontrol=1
  \def\@bibitem#1{%
    \st@bibitem{#1}\a@@label{#1}\ignorespaces}
  \def\@lbibitem[#1]#2{%
    \st@lbibitem[#1]{#2}\a@@label{#2}\ignorespaces}
  \def\a@@label#1{%
    \gdef\a@lab{\smash{\normalfont\small#1}}
    \ifvmode
      \if@inlabel
        \global\setbox\@labels\hbox{%
          \llap{\a@lab\let\a@lab\relax
                \kern\@totalleftmargin\kern\marginparsep}%
          \box\@labels}%
      \fi
    \fi}
\fi
\documentclass[12pt,letterpaper]{article}
\usepackage{amsmath,amssymb,array,calc,rotating,epsfig,psfrag}
\usepackage{hyperref}
\ifnum\draftcontrol=1
\fi
\renewcommand\baselinestretch{1.25}
\renewcommand\section{\@startsection {section}{1}{\z@}%
                                   {-3.5ex \@plus -1ex \@minus -.2ex}%
                                   {2.3ex \@plus.2ex}%
                                   {\normalfont\large\bfseries}}
\renewcommand\subsection{\@startsection{subsection}{2}{\z@}%
                                     {-3.25ex\@plus -1ex \@minus -.2ex}%
                                     {1.5ex \@plus .2ex}%
                                     {\normalfont\normalsize\bfseries}}
\renewcommand\subsubsection{\@startsection{subsubsection}{3}{\z@}%
                                     {-3.25ex\@plus -1ex \@minus -.2ex}%
                                     {1.5ex \@plus .2ex}%
                                     {\normalfont\normalsize\it}}

\numberwithin{equation}{section}

\def\projective   {{\mathbb P}}

\def\reals        {{\mathbb R}}

\def\revise#1       {\marginpar{\rule{2mm}{1cm} #1}}

\def\RR{\reals}
\def\PP{\projective}
\def\RP{\RR\PP}

\def\R{{\rm R}}

\def\sqr#1#2{{\vcenter{\vbox{\hrule height.#2pt  
 \hbox{\vrule width.#2pt height#1pt \kern#1pt
 \vrule width.#2pt}\hrule height.#2pt}}}}

\def\yboxit#1#2{\vbox{\hrule height #1 \hbox{\vrule width #1
\vbox{#2}\vrule width #1 }\hrule height #1 }}
\def\fillbox#1{\hbox to #1{\vbox to #1{\vfil}\hfil}}
\def\ybox{{\lower 1.3pt \yboxit{0.4pt}{\fillbox{8pt}}\hskip-0.2pt}}

\def\comments#1{}

\def\half{{\frac12}}

\def\Re{{\rm Re\hskip0.1em}}
\def\Im{{\rm Im\hskip0.1em}}

\def\ket#1{|#1\rangle}

\def\CD{{\cal D}}

\def\CN{{\cal N}}

\def\P{\BP}

\def\II{\relax{I\kern-.10em I}}

\def\IZ{\relax\ifmmode\mathchoice
{\hbox{\cmss Z\kern-.4em Z}}{\hbox{\cmss Z\kern-.4em Z}}
{\lower.9pt\hbox{\cmsss Z\kern-.4em Z}}
{\lower1.2pt\hbox{\cmsss Z\kern-.4em Z}}\else{\cmss Z\kern-.4em
Z}\fi}
\def\IB{\relax{\rm I\kern-.18em B}}
\def\IC{{\relax\hbox{$\inbar\kern-.3em{\rm C}$}}}
\def\ID{\relax{\rm I\kern-.18em D}}
\def\IE{\relax{\rm I\kern-.18em E}}
\def\IF{\relax{\rm I\kern-.18em F}}
\def\IG{\relax\hbox{$\inbar\kern-.3em{\rm G}$}}
\def\IGa{\relax\hbox{${\rm I}\kern-.18em\Gamma$}}
\def\IH{\relax{\rm I\kern-.18em H}}
\def\II{\relax{\rm I\kern-.18em I}}
\def\IK{\relax{\rm I\kern-.18em K}}
\def\IP{\relax{\rm I\kern-.18em P}}
\def\inbar{\,\vrule height1.5ex width.4pt depth0pt}

\font\cmss=cmss10 \font\cmsss=cmss10 at 7pt
\def\IR{\relax{\rm I\kern-.18em R}}

\def\BR{\IR}
\def\BP{\IP}
\def\BR{\IR}

\def\Bid{{\mathchoice {\rm {1\mskip-4.5mu l}} {\rm
{1\mskip-4.5mu l}} {\rm {1\mskip-3.8mu l}} {\rm {1\mskip-4.3mu l}}}}

\def\lp10{l_P^{10}}
\def\lp11{l_P^{11}}
\newcommand{\nc}{\newcommand}
\nc{\rnc}{\renewcommand}
\nc{\CY}{Calabi-Yau}
\nc{\CYM}{Calabi-Yau manifold}
\nc{\CYMs}{Calabi-Yau manifolds}
\nc{\DB}{D-Brane}
\nc{\DBs}{D-Branes}
\nc{\SUSY}{supersymmetry}
\nc{\Kah}{K\"ahler}
\nc{\cs}{complex structure}
\nc{\beq}{\begin{equation}}
\nc{\eeq}{\end{equation}}
\nc{\beqa}{\begin{eqnarray}}
\nc{\eeqa}{\end{eqnarray}}
\nc{\ntwo}{${\cal N}=2$}
\nc{\nOne}{${\cal N}=1$}
\nc{\hs}{\hspace{0.2in}}
\nc{\Z}{{\mathbb Z}}
\rnc{\P}{{\mathbb P}}
\rnc{\RP}{{\mathbb {RP}}}
\nc{\WP}{\mathbb{WP}}
\nc{\slag}{special Lagrangian}
\nc{\cn}{\C^n}
\nc{\rn}{\R^n}
\def\ket#1{|#1\rangle}

\nc{\SO}{SO}
\nc{\Sp}{Sp}
\nc{\SU}{SU}

\nc{\Wtree}{W_{\mathrm tree}}
\nc{\Weff}{W_{\mathrm eff}}

\catcode`\@=12

\begin{document}

\title{\Large \bf Bubbling Defect CFT's\\[1cm]}

\pubnum{%
hep-th/0604155}
\date{April 2006}

\author{Jaume Gomis\footnote{\tt{jgomis@perimeterinstitute.ca}}\; and
Christian R\"omelsberger\footnote{\tt{cromelsberger@perimeterinstitute.ca}}\\[1cm]
\it Perimeter Institute \\
\it 31 Caroline St. N. \\
\it Waterloo, ON N2L 2Y5 , Canada \\[1cm]
}

\Abstract{\\We study the gravitational description of conformal half-BPS 
domain wall operators in $\CN=4$ SYM, which are described by defect CFT's. 
These defect CFT's arise in the low energy limit of a Hanany-Witten like brane setup and are described in a probe brane approximation by a Karch-Randall brane configuration.
The gravitational backreaction takes the five-branes in
$AdS_5\times S^5$ through a geometric transition and turns them into 
appropriate
fluxes which are supported on non-trivial three-spheres.\\[1cm]}

\enlargethispage{1.5cm}
\makepapertitle
\vfill \eject 
\tableofcontents

\body

\version\versionno

\section{Introduction}

Local gauge invariant  operators are labeled by a point in spacetime. Such operators  can be
constructed by combining in a gauge invariant way the fields appearing in the action. This is, however, not the only way to define local operators. Operators which cannot be written in a local way in terms of the fields appearing in the action are quite ubiquitous in quantum field theory. Examples of this class of operators include twist operators in conformal field theory and  ``soliton" operators in gauge theories (see e.g. \cite{Kapustin:2005py} for a recent discussion).  The insertion of an
 operator in the path integral   has the effect of introducing  at the location of the operator a very specific singularity  for the fields that appear in the action.

There is no essential  limitation in quantum field theory  restricting the class of admissible singularities of the fundamental fields to be point-like; in principle they can be defined on any  defect in spacetime. In four dimensional field theories one may consider line, surface and domain wall operators on top of the more familiar local operators labeled by a point in spacetime. In favorable circumstances -- e.g. Wilson loops -- these operators can be written down using the fields appearing in the action while in others -- e.g. 't~Hooft loops -- the operators are defined by the singularity they produce for the fundamental fields in the action at the location of the defect.

A convenient way to construct a  defect operator is to introduce additional degrees of freedom localized on the defect. The extra degrees of freedom encode the type of singularity
produced by the defect operator. The study of defect operators can then by mapped to the problem of 
studying the defect field theory describing the coupling of a four dimensional field theory to the degrees of freedom living on the defect. Demanding invariance of the defect operator under some symmetry constraints the geometry of the defect as well as the allowed degrees of freedom that can be added to the defect. In particular, demanding invariance under the conformal group on the defect leads to defect conformal field theories \cite{Cardy:1984bb,McAvity:1995zd}.

The AdS/CFT conjecture \cite{Maldacena:1997re,Gubser:1998bc,Witten:1998qj} requires that all gauge invariant operators in ${\cal N}=4$ SYM have a  realization in the bulk description.  This program has been successfully carried out for the  half-BPS local operators in ${\cal N}=4$ SYM \cite{Witten:1998qj,Aharony:1999ti}, where the operators can be identified with D-branes in the bulk
\cite{Corley:2001zk,Berenstein:2004kk}. Recently, the dictionary has been enlarged  \cite{Gomis:2006sb} (see also \cite{Drukker:2005kx,Yamaguchi:2006tq}) to include all the half-BPS Wilson loop operators in ${\cal N}=4$ SYM, which have also been identified with D-branes in the bulk. The half-BPS Wilson loop operators are constructed \cite{Gomis:2006sb} by integrating out in the defect conformal field theory  the localized degrees of freedom   living on the loop  that are introduced by the bulk D-branes.

In this paper we study half-BPS domain wall operators in ${\cal N}=4$ SYM and their corresponding bulk description. Domain wall operators can be identified with a defect conformal field theory describing the coupling of ${\cal N}=4$ SYM
to additional degrees of freedom localized in an  $\BR^{1,2}\subset \BR^{1,3}$ defect. Supersymmetry requires that  the degrees of freedom living on the defect fill three dimensional hypermultiplets. These new localized hypermultiplet degrees of freedom arise from the  presence of branes (D5 and NS5-branes) in $AdS_5\times S^5$ that end on a common $\BR^{1,2}$ defect at the boundary. For each half-BPS configuration of five-branes in $AdS_5\times S^5$, we may associate a defect conformal field theory or equivalently a half-BPS domain wall operator.

 The defect conformal field theory can be derived by studying the low energy effective field 
theory\footnote{This brane configuration is a generalization of   the brane construction in 
\cite{Hanany:1996ie} studied in the context of three dimensional mirror symmetry.}
 on $N$ D3-branes in the presence of D5 and NS5-branes intersecting  the D3-branes along an 
$\BR^{1,2}\subset \BR^{1,3}$ defect.  The data which determines the defect conformal field theory 
under study is the number of D3-branes which end on each of the D5-branes and NS5-branes.  
 In the decoupling/near horizon limit the five-branes span a family of 
$AdS_4\times\tilde{S}^2$ and   $AdS_4\times S^2$ geometries  in  $AdS_5\times S^5$ respectively,  of the type found in 
\cite{Karch:2000ct}.  Each such array of five-branes in the bulk corresponds to a half-BPS domain 
wall operator.

 We study the backreaction on the $AdS_5\times S^5$ background due to the configuration of five-branes 
dual to a specific domain wall operator. We show that the solution of the supergravity BPS equations is
determined by specifying boundary conditions on a two dimensional surface in the ten dimensional geometry. 
These boundary conditions encode the location where either an $S^2$ or an ${\tilde S }^2$ shrinks to zero 
size in a smooth manner.  This result generalizes the work of LLM \cite{Lin:2004nb} -- which applies to 
the half-BPS local operators --  to the geometries dual to half-BPS domain wall\footnote{Recently, 
Yamaguchi \cite{Yamaguchi:2006te} has made an analogous ansatz relevant for Wilson loops.} operators. 
 
 The $AdS_5\times S^5$ vacuum solution corresponds to an infinite strip, where at the bottom of the strip 
${\tilde S }^2$ shrinks to zero size while at the top of the strip $S^2$ shrinks to zero size. In the probe 
approximation, a  D5-brane is located on the top boundary of the strip while a NS5-brane is located at the 
bottom. The precise location of brane is determined by the amount of D3-brane charge carried by the 
five-branes.
 
In the backreacted geometry we expect these five-branes to be replaced by bubbles of flux
\cite{Gopakumar:1998vy}\cite{Klebanov:2000hb}\cite{Maldacena:2000yy}\cite{Chamseddine:1997nm}\cite{Chamseddine:1997mc}, i.e. by smooth non-contractible three-cycles 
supporting the appropriate amount of three-form flux. There are two ways  for this to happen.
In one the geometry stays finite and there appears a change in the coloring of the boundary. For
a D5-brane this corresponds to having a finite size $S^2$ and $\tilde S^2$ shrinking in a finite
segment of the upper boundary. The second possibility is an infinite throat developing on the
upper boundary with the $S^2$ shrunk on both boundaries of the throat. In both cases there appears 
a smooth three-sphere which can support the three-form flux.

It would be very interesting to get a better understanding of the behavior of the solutions near 
the defects, how the boundary conditions work and how the fluxes, brane charges and changes in the
rank of the gauge group are related to the boundary conditions as we understand them.

The plan of the rest of the paper is as follows. In section \ref{secdomainwalls} we study the 
brane configuration whose low energy effective field theory yields the defect conformal field 
theories we are interested in. In section \ref{secprobe} we consider the bulk description of the 
half-BPS domain wall operators in the probe approximation. In section \ref{secsugra} we 
derive the BPS equations and some important normalization conditions for spinor bilinears by
realizing the (super) symmetry algebra in type IIB supergravity. In section \ref{secwallgeometries}
we first discuss the $AdS_5\times S^5$ solution of the BPS equations, then we 'bootstrap' the 
general BPS equations to get a second order PDE for one remaining spinor variable and finally we 
discuss the supersymmetries of probe branes, boundary conditions and the general structure of 
solutions.

While this paper was in preparation, a paper \cite{Lunin:2006xr} appeared which overlaps with ours. 

\section{Half-BPS Domain Wall Operators and Defect Field Theory}\label{secdomainwalls}

Defect operators can be defined by introducing degrees of freedom localized on the defect. The theory 
that captures the interactions of the localized degrees of freedom with those of ${\cal N}=4$ SYM is a defect conformal field theory if we impose that the defect preserves conformal invariance.

The description of defect operators in terms of defect conformal field theories naturally suggests the construction of such theories as low energy limits of branes in string theory. The strategy is to consider 
brane configurations involving D3-branes  together with other branes intersecting the D3-branes along a defect. The low energy effective field theory is described by ${\cal N}=4$ SYM coupled to the degrees of freedom localized on the defect introduced by the other branes. 

Here we are interested in half-BPS domain wall operators in ${\cal N}=4$ SYM. These operators arise by considering the following brane configuration:
\beq\begin{tabular}{|l|c|c|c|c|c|c|c|c|c|c|}
\hline
&0&1&2&3&4&5&6&7&8&9\\\hline
D3&x&x&x&x&&&&&&\\\hline
D5&x&x&x&&x&x&x&&&\\\hline
NS&x&x&x&&&&&x&x&x\\
\hline
\end{tabular}\eeq
The degrees of freedom associated with  this brane configuration is the ${\cal N}=4$ SYM multiplet together 
hypermultiplets \cite{Hanany:1996ie} localized on $\BR^{1,2}\subset \BR^{1,3}$. 

\begin{figure}[bth]
\centerline{ \epsfig{file=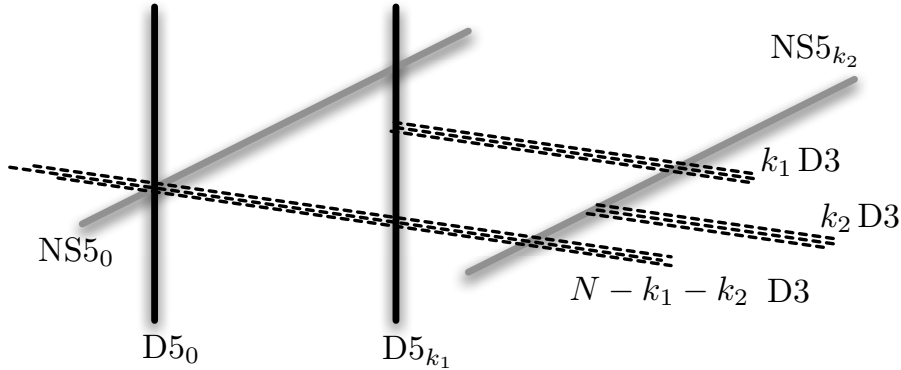,width=12cm}}
\caption{\sl The brane configuration in flat space.}
\end{figure}

The action for this defect field theory in the absence of NS5-branes  and when none of the D3-branes end on the D5-branes has been constructed in \cite{Karch:2000gx,DeWolfe:2001pq,Erdmenger:2002ex}. We refer the reader to these references for the detailed form of the action.

In order to obtain more general domain wall operators, one can allow for configurations where a number of D3-branes end on a D5-brane or NS5-brane. We label by D5$_k$/NS5$_k$ a D5/NS5-brane where $k$ D3-branes end. Each choice of partitioning the $N$ D3-branes among the five-branes corresponds to a different half-BPS domain wall operator. This is similar to the construction of half-BPS Wilson loops \cite{Gomis:2006sb} using D-branes. 

The D3-branes ending on the five-branes have the effect of introducing magnetic charge on the five-brane worlvolume. It would be very interesting to construct explicitly this general class of defect field theories and study in detail the singularities produced on the ${\cal N}=4$ SYM fields by the corresponding half-BPS domain wall operator.

We conclude this section with the analysis of the symmetries of these defect conformal field theories, which play a crucial role in the following sections, where the supergravity description of these operators is studied.

The bosonic symmetry algebra is 
$SO(2,3)\times SO(3)\times SO(3)=Sp(4,\BR)\otimes SO(4)$, which is generated 
by generators $M^A$ and $M^L$. In the $(4,4)$ those generators $(M^A)^\alpha{}_\beta$ 
and $(M^L)^{\dot\alpha}{}_{\dot\beta}$ can 
be chosen real and the supersymmetry generators\footnote{The undotted index 
$\alpha=1,\cdots,4$ is an index in the real fundamental representation 
of $Sp(4,\BR)$, whereas the dotted index $\dot\alpha=1,\cdots,4$ is an  
index in the real fundamental representation of $SO(4)$.}
$Q_{\alpha\dot\alpha}$ are Hermitean
\beq
Q_{\alpha\dot\alpha}^\dagger=Q_{\alpha\dot\alpha},
\eeq
they transform as
\beq
[M^A,Q_{\alpha\dot\alpha}]=
-(M^A)^\beta{}_\alpha Q_{\beta\dot\alpha}\qquad{\rm and}\qquad
[M^L,Q_{\alpha\dot\alpha}]=
-(M^L)^{\dot\beta}{}_{\dot\alpha}Q_{\alpha\dot\beta}.
\eeq
Their anti commutation relation is
\beq
\{Q_{\alpha\dot\alpha},Q_{\beta\dot\beta}\}=
i(M_AJ)_{\alpha\beta}I_{\dot\alpha\dot\beta}M^A-
iJ_{\alpha\beta}(M_LI)_{\dot\alpha\dot\beta}M^L,
\eeq
where $J_{\alpha\beta}$ is the real, invariant, antisymmetric matrix of $Sp(4,\BR)$ and 
$I_{\dot\alpha\dot\beta}$ is the real, invariant, symmetric matrix of $SO(4)$. Therefore, these defect conformal field theories are invariant under an $OSp(4|4)$ subalgebra of the $SU(2,2|4)$ algebra of ${\cal N}=4$ SYM.

The supersymmetry generators can be contracted with real Grassmann variables
$\hat\epsilon_{\alpha\dot\alpha}$ to form Hermitean generators
\beq
\hat\epsilon^{\alpha\dot\alpha} Q_{\alpha\dot\alpha}=
\hat\epsilon^{\alpha\dot\alpha}{}^\ast Q^\dagger_{\alpha\dot\alpha}.
\eeq

\section{Probe Branes in $AdS_5\times S^5$}\label{secprobe}

In this section we study in the probe approximation the branes in $AdS_5\times S^5$ which correspond to the half-BPS domain wall operators described in the previous section. In the next sections we study the backreaction produced by these branes and find the equations which determine the asymptotically AdS geometries dual to the defect operators. 

In order to make manifest the $SO(2,3)\otimes SO(3)\otimes SO(3)$ symmetry of the half-BPS domain wall operators we foliate the $AdS_5$ geometry by $AdS_4$ slicings
\beq
ds^2=R^2\left(\cosh^2(x)ds_{AdS_4}+dx^2\right),
\eeq
where $R$ is the radius of curvature of $AdS_5$ and $S^5$. We also foliate $S^5$ by $S^2\times\tilde{S}^2$ slicings
\beq
ds^2=R^2\left(dy^2+\cos^2(y)d\Omega_2+\sin^2(y)d\tilde{\Omega}_2\right),
\eeq
where $d\Omega_2 (d\tilde{\Omega}_2)$ is the metric on a unit $S^2$ ($\tilde{S}^2$).

We note that in this parametrization the $AdS_5\times S^5$ metric can be represented by an $AdS_4\times S^2\times\tilde{S}^2$ fibration over a strip, whose length is parametrized by $x$ and width by $y$.  At the $y=0$ boundary of the strip $\tilde{S}^2$ shrinks smoothly to zero size while at the $y=\pi/2$ boundary $S^2$ shrinks smoothly.

The D5$_k$-brane in the previous section becomes in the near horizon limit a D5-brane in $AdS_5\times S^5$ with an $AdS_4\times\tilde{S}^2$ worldvolume and with $k$ units of magnetic flux dissolved on the D5-brane. The details of this solution can be found in \cite{Karch:2000gx} and the analysis of supersymmetry in \cite{Skenderis:2002vf}. A D5$_k$-brane sits at $y=\pi/2$ and at $x(k)=\sinh^{-1}(\pi k/R)$. 

Similarly, the solution for the NS5$_k$-brane can also be found. It spans an $AdS_4\times S^2$ geometry in $AdS_5\times S^5$ and has $k$ units of magnetic flux dissolved in it. Now, the NS5$_k$-brane sits at $y=0$ and at $x(k)=\sinh^{-1}(\pi k/R)$. 

We will show in section \ref{secprobebc} that all these five-branes preserve exactly the same supersymmetries and coincide with the supersymmetries preserved by the defect conformal field theory.

We therefore see that any half-BPS domain wall operator in the probe approximation can be 
characterized by a collection of points on the appropriate boundary of the strip which characterizes 
$AdS_5\times S^5$. To each D5$_k$-brane of the microscopic description of the defect conformal 
field theory we associate a point at the $y=\pi/2$ boundary of the strip located at $x(k)$, 
where $k$ is the number of D3-branes ending on the D5-brane. Similarly, to each NS5$_k$-brane 
of the microscopic description of the defect conformal field theory we 
associate a point at the $y=0$ boundary of the strip located at $x(k)$, where $k$ is the 
number of D3-branes ending on the NS5-brane. Therefore, to a given half-BPS domain wall 
operator we can associate the following strip
\begin{figure}[bth]
\centerline{ \epsfig{file=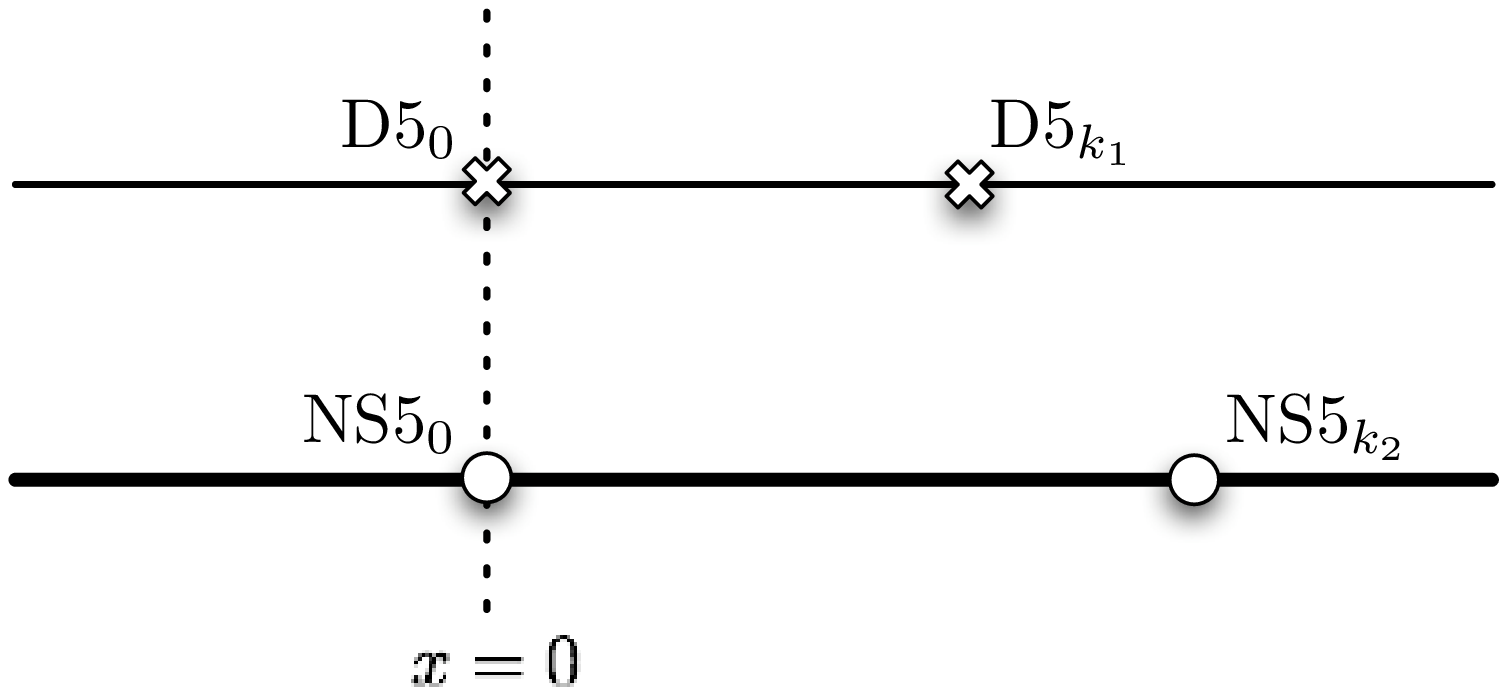,width=12cm}}
\caption{\sl Probe branes in $AdS_5\times S^5$}
\end{figure}

The goal of the rest of the paper is to find the BPS equations in Type IIB supergravity which 
determine the backreaction\footnote{In \cite{Fayyazuddin:2002bm} the supergravity equations for intersecting D3/D5 branes were analyzed.} produced by a collection of five-branes corresponding dual to a defect 
conformal field theory.

\section{The supergravity solution}\label{secsugra}

In this section we derive the BPS equations by reproducing the $OSp(4|4)$ supersymmetry algebra
of the half-BPS domain wall operators in 
Type IIB
supergravity. This consists of two parts, the invariance of the background under the 
(super) symmetry transformations as well as the closure of the (super) symmetry algebra.
This section is very technical. Readers who do not want to go into technical details
can just read \ref{secansatz}. The result of this section is the BPS equations
(\ref{bpsdv}), (\ref{bpsgv1}) and (\ref{bpsgv2}) together with the normalization conditions
(\ref{normcond}). We are using the type IIB supergravity
conventions of \cite{Schwarz:1983qr,Gauntlett:2005ww} with a mostly + signature. The
gamma matrix conventions are summarized in Appendix \ref{appclifford}.

\subsection{The Ansatz}\label{secansatz}

The bosonic symmetry group is $SO(2,3)\times SO(3)\times SO(3)$ and the
10 dimensional space time is a $AdS_4\times S^2\times S^2$ fibration
over a two dimensional base space $M_2$. The most general vielbein ansatz 
is
\beq\begin{array}{l}
e^\mu=A_1\,e^{\hat \mu},\qquad \mu=0,1,2,3,\\
e^m=A_2\,e^{\tilde m}, \qquad m=4,5,\\
e^i=A_3\,e^{\check i}, \qquad i=6,7, \\
e^a, \qquad\qquad\qquad  a=8,9,
\end{array}\eeq
where $e^{\hat \mu}$ is a vielbein on a unit $AdS_4$, $e^{\tilde m}$ 
and $e^i$ are vielbeins on the unit $S^2$ and $\tilde{S}^2$ respectively and $e^a$ is a vielbein on
$M_2$.
The most general self dual 5-form flux has the form
\beq
F=f_a(e^{0123a}+\epsilon_{ab}\,e^{4567b}),
\eeq
where $f_a\,e^a$ is a real 2-form on $M_2$. The most general dilaton-axion $P$
and 3-form fluxes $G$ are given in terms of the 
complex 1-forms $p^{(4)}=p_a\,e^a$, $g^{(4)}=g_a\,e^a$ and $h^{(4)}=h_a\,e^a$ 
on $M_2$
\beq
P=p_a\,e^a\qquad{\rm and}\qquad
G=g_a\,e^{45a}+ih_a\,e^{67a}.
\eeq
The most general $U(1)$-R connection is given by the two dimensional
connection $q^{(4)}=q_a\,e^a$ on $M_2$
\beq
Q=q_a\,e^a.
\eeq

\subsection{A specific basis for $OSp(4|4)$}

We will derive the BPS equations by heavily using the explicit form
of the symmetry algebra $OSp(4|4)$ in terms of the clifford algebras
of $SO(2,3)\times SO(3)\times SO(3)$.
%For later it is useful to make the map from $Sp(4,\BR)\times SO(4)$ to
%$SO(2,3)\times SO(3)\times SO(3)$ more precise. 
We use the Clifford 
algebra conventions of Appendix \ref{appclifford}. A basis of generators
in the 4 of $Sp(4,\BR)$ is
\beq\label{rotmatrix1}
M^{\mu\nu}\sim\half\gamma^{\mu\nu},\qquad
M^\mu\sim\half\gamma^\mu.
\eeq
The matrix
\beq
J=iD^{(1)}\gamma^{(1)}
\eeq
is invariant and antisymmetric. One can coose a basis of Majorana
spinors
\beq
\chi_\alpha^\ast=B^{(1)}\chi_\alpha
\eeq
and a dual basis of spinors $\chi^\alpha$ such that
\beq
\chi^\alpha{}^t\chi_\beta=\delta^\alpha_\beta.
\eeq
Then 
\beq\label{rotmatrix2}
(M^{\mu\nu})^\alpha{}_\beta=\chi^\alpha{}^tM^{\mu\nu}\chi_\beta
\qquad{\rm and}\qquad 
(M^\mu)^\alpha{}_\beta=\chi^\alpha{}^tM^\mu\chi_\beta
\eeq
are real and 
\beq
J_{\alpha\beta}=\chi_\alpha^tJ\chi_\beta
\eeq
is real, invariant and antisymmetric.

Similarly,
\beq\begin{array}{ll}
M^{mn}\sim\half(\gamma^{mn}\otimes\Bid),\qquad&
M^m\sim\frac{i}{2}(\gamma^m\otimes\Bid),\\
M^{ij}\sim\half(\Bid\otimes\gamma^{ij}),\qquad&
M^i\sim\frac{i}{2}(\Bid\otimes\gamma^i)
\end{array}\eeq
are a basis of generators of the 4 of $SO(4)$, the matrix 
\beq
I=D^{(2)}\otimes D^{(3)}
\eeq
is invariant and symmetric. One can coose a basis of Majorana
spinors
\beq
\chi_{\dot\alpha}^\ast=(B^{(2)}\otimes B^{(3)})\chi_{\dot\alpha}
\eeq
and a dual basis of spinors $\chi^{\dot\alpha}$ such that
\beq
\chi^{\dot\alpha}{}^t\chi_{\dot\beta}=\delta^{\dot\alpha}_{\dot\beta}.
\eeq
Then $(M^{mn})^{\dot\alpha}{}_{\dot\beta}$,$(M^m)^{\dot\alpha}{}_{\dot\beta}$, 
$(M^{ij})^{\dot\alpha}{}_{\dot\beta}$ and $(M^i)^{\dot\alpha}{}_{\dot\beta}$ are
real and $I_{\dot\alpha\dot\beta}$ is real, invariant and symmetric.

The anticommutation relation of the supercharges is then
\beq\label{anticommutator}\begin{array}{rl}
\{Q_{\alpha\dot\alpha},Q_{\beta\dot\beta}\}=&
\frac{1}{4}(\bar\chi_{\alpha\dot\alpha}(\gamma^{(1)}\gamma_{\mu\nu}\otimes\Bid\otimes\Bid)\chi_{\beta\dot\beta})M^{\mu\nu}+
\frac{1}{2}(\bar\chi_{\alpha\dot\alpha}(\gamma^{(1)}\gamma_\mu\otimes\Bid\otimes\Bid)\chi_{\beta\dot\beta})M^\mu-\\
&\frac{1}{4}(\bar\chi_{\alpha\dot\alpha}(\gamma^{(1)}\otimes\gamma_{mn}\otimes\Bid)\chi_{\beta\dot\beta})M^{mn}-
\frac{i}{2}(\bar\chi_{\alpha\dot\alpha}(\gamma^{(1)}\otimes\gamma_m\otimes\Bid)\chi_{\beta\dot\beta})M^m-\\
&\frac{1}{4}(\bar\chi_{\alpha\dot\alpha}(\gamma^{(1)}\otimes\Bid\otimes\gamma_{ij})\chi_{\beta\dot\beta})M^{ij}-
\frac{i}{2}(\bar\chi_{\alpha\dot\alpha}(\gamma^{(1)}\otimes\Bid\otimes\gamma_i)\chi_{\beta\dot\beta})M^i.
\end{array}\eeq

\subsection{Symmetries and the Killing spinor equations}

The Killing spinors $\epsilon$ have to transform in the $(4,2,2)$ 
representation of the Bosonic symmetry group $SO(2,3)\times SO(3)\times SO(3)$.
The Bosonic symmetries are realized by Killing vector fields. Those act 
through the Lie derivative on the Killing spinors.

For a given point $Q$ on $M_{10}$ there is a 
$SO(1,3)\times SO(2)\times SO(2)$ stabilizer group. The Lie derivative
for this stabilizer group acts by rotations 
\beq\label{stabilizeraction}
\half\gamma^{\mu\nu},\qquad 
\half\gamma^{mn}\qquad{\rm and}\qquad
\half\gamma^{ij}
\eeq
on the Killing spinor $\epsilon$ at $Q$. For the tangent vectors 
$e_{\hat\mu}$, $e_{\tilde m}$ and $e_{\check i}$ at $Q$ 
there are unique Killing vector fields which generate a geodesic 
through $Q$ in the fiber. The Lie derivative along those Killing vector 
fields at $Q$ are given by the covariant derivatives
\beq
\pounds_{\hat\mu}\epsilon=\nabla_{\hat\mu}\epsilon,\qquad
\pounds_{\tilde m}\epsilon=\nabla_{\tilde m}\epsilon\qquad{\rm and}\qquad
\pounds_{\check i}\epsilon=\nabla_{\check i}\epsilon.
\eeq

The fact that the Killing spinors $\epsilon$ have to transform in 
the $(4,2,2)$ representation of the Bosonic symmetry group 
$SO(2,3)\times SO(3)\times SO(3)$ implies that
the Lie derivative action on a Killing spinor can be reproduced by a 
matrix action $N_\mu$, $N_m$ and $N_i$ which is consistent with 
(\ref{stabilizeraction}), (\ref{rotmatrix1}) and (\ref{rotmatrix2}). Consistency and 
the 10-dimensional chirality condition then imply that
\beq\label{definen}\begin{array}{l}
N_\mu=\half n(\gamma_\mu\otimes\Bid\otimes\Bid\otimes\gamma_8) n^{-1},\\
N_m=\frac{i}{2} n(\Bid\otimes\gamma_m\otimes\Bid\otimes\gamma_8) n^{-1}\qquad
{\rm and}\\
N_i=\frac{i}{2} n(\Bid\otimes\Bid\otimes\gamma_i\otimes\gamma_8) n^{-1},
\end{array}\eeq
where $n$ is a unitary matrix of the form
\beq
n=
%f_{\eta_1\eta_2\eta_3\eta_4}\gamma^{(\eta_1\eta_2\eta_3\eta_4)}=
f_{\eta_1\eta_2\eta_3\eta_4}(\gamma^{(1)})^{\frac{2-\eta_1}{2}}
(\gamma^{(2)})^{\frac{2-\eta_2}{2}}(\gamma^{(3)})^{\frac{2-\eta_3}{2}}
(\gamma^{(4)})^{\frac{2-\eta_4}{2}}.
\eeq
Note that (\ref{definen}) does not fix $n$ uniquely.
The Killing spinor equations then have the form
\beq\label{killingeq}\begin{array}{c}
(\nabla_{\hat\mu}-N_\mu)\epsilon=0,\\
(\nabla_{\tilde m}-N_m)\epsilon=0,\\
(\nabla_{\check i}-N_i)\epsilon=0.
\end{array}\eeq

To solve those 10-dimensional Killing spinor equations let us first have 
a look at the simplified 8-dimensional and 2-dimensional Killing spinor 
equations
\beq\begin{array}{c}
(\nabla_{\hat\mu}-\frac{\eta_1}{2}(\gamma_\mu\otimes\Bid\otimes\Bid))
\chi^{(\eta_1,\eta_2,\eta_3)}_{\alpha\dot\alpha}=0,\\
(\nabla_{\tilde m}-\frac{i\eta_2}{2}(\Bid\otimes\gamma_m\otimes\Bid))
\chi^{(\eta_1,\eta_2,\eta_3)}_{\alpha\dot\alpha}=0,\\
(\nabla_{\check i}-\frac{i\eta_3}{2}(\Bid\otimes\Bid\otimes\gamma_i))
\chi^{(\eta_1,\eta_2,\eta_3)}_{\alpha\dot\alpha}=0,\\
(\Bid-\eta_4\gamma^8)\zeta^{(\eta_4)}=0.
\end{array}\eeq
The solutions to those equations transform in the $(4,2,2)$
of $SO(1,3)\times SO(2)\times SO(2)$. This representation allows for 
a reality condition. A basis of the real representation is labelled 
by $(\alpha,\dot\alpha)$
\beq
(B^{(1)}\otimes B^{(2)}\otimes B^{(3)})^{-1}
(\chi^{(1,1,1)}_{\alpha\dot\alpha})^\ast=
\chi^{(1,1,1)}_{\alpha\dot\alpha}.
\eeq

Define
\beq
\epsilon_0=
\hat\epsilon^{\alpha\dot\alpha}\chi^{(1,1,1)}_{\alpha\dot\alpha}\otimes
\zeta^{(1)},
\eeq
where $\hat\epsilon^{\alpha\dot\alpha}$ is a real Grassman variable defined in the 
last section. It is not hard to see that
\beq\label{killingspinor}
\epsilon=
n\epsilon_0=
\hat\epsilon^{\alpha\dot\alpha}f_{\eta_1\eta_2\eta_3\eta_4}
\chi^{(\eta_1,\eta_2,\eta_3)}_{\alpha\dot\alpha}\otimes\zeta^{(\eta_4)}=
\hat\epsilon^{\alpha\dot\alpha}
\chi^{(\eta_1,\eta_2,\eta_3)}_{\alpha\dot\alpha}\otimes
\zeta_{\eta_1\eta_2\eta_3}
\eeq
is the general solution of (\ref{killingeq}).

The 10-dimensional chirality condition imposes
\beq\label{chiralitycond}
\gamma^{(4)}\zeta_{-\eta}=\zeta_\eta,
\eeq
whereas the reality condition implies
\beq
\ast\epsilon=-\eta_1\eta_2\eta_3\hat\epsilon^{\alpha\dot\alpha}
\chi^{(\eta_1,\eta_2,\eta_3)}_{\alpha\dot\alpha}\otimes
(\ast\zeta_{-\eta_1,-\eta_2,\eta_3}).
\eeq
This allows to translate the dilatino and gravitino variation equations 
into equations which depend linearly on 
\beq
\hat\epsilon^{\alpha\dot\alpha}\chi^{(\eta_1,\eta_2,\eta_3)}_{\alpha\dot\alpha}.
\eeq
%{\bf(Can we strip off such factors from the BPS equations?)}

\subsection{The Dilatino and Gravitino variation equations}

The dilatino variation equation
\beq
P_M\,\gamma^M\ast\epsilon+\frac{1}{24}\,G_{MNP}\gamma^{MNP}\epsilon=0
\eeq
turns into\footnote{Note that $\ast\zeta=\gamma^8\sigma^{(2,2,2)}\zeta^\ast$ is the covariant
complex conjugation.}
\beq\label{dv}
ip_a\sigma^{(1,1,0)}\gamma^a\ast\zeta-
\frac{ig_a}{24}\sigma^{(1,0,1)}\gamma^a\zeta+
\frac{h_a}{24}\sigma^{(1,1,0)}\gamma^a\zeta=0.
\eeq

To calculate the gravitino variation equations
\beq
\CD_M\epsilon+\frac{i}{480}F_{PQRST}\gamma^{PQRST}\gamma_M\epsilon
-\frac{1}{96}G_{PQR}(\gamma_M{}^{PQR}-9\delta^P_M\gamma^{QR})\ast\epsilon=0,
\eeq 
we need the spin connection
\beq\begin{array}{l}
\omega_{\mu\nu\rho}=\frac{1}{A_1}\omega_{\hat\mu\hat\nu\hat\rho}, \qquad
\omega_{\mu\nu a}=\eta_{\mu\nu}\frac{\partial_a A_1}{A_1}, \\
\omega_{mnp}=\frac{1}{A_2}\omega_{\tilde m\tilde n\tilde p}, \qquad
\omega_{mna}=\delta_{mn}\frac{\partial_a A_2}{A_2}, \\
\omega_{ijk}=\frac{1}{A_3}\omega_{\check i\check j\check k}, \qquad
\omega_{ija}=\delta_{ij}\frac{\partial_a A_3}{A_3}, \\
\omega_{aab}.
\end{array}\eeq
The gravitino variation equations turn into
\beqa\nonumber
\frac{i}{2A_1}\sigma^{(2,1,1)}\zeta+
\frac{\partial_aA_1}{2A_1}\gamma^a\zeta
+\frac{f_a}{240}\gamma^a\sigma^{(1,0,0)}\zeta
-\frac{g_a}{96}\gamma^a\sigma^{(0,1,1)}\ast\zeta
-\frac{ih_a}{96}\gamma^a\ast\zeta=0,\\\nonumber
-\frac{1}{2A_2}\sigma^{(0,2,1)}\zeta+
\frac{\partial_aA_2}{2A_2}\gamma^a\zeta
-\frac{f_a}{240}\gamma^a\sigma^{(1,0,0)}\zeta
+\frac{g_a}{32}\gamma^a\sigma^{(0,1,1)}\ast\zeta
-\frac{ih_a}{96}\gamma^a\ast\zeta=0,\\\nonumber
-\frac{1}{2A_3}\sigma^{(0,0,2)}\zeta+
\frac{\partial_aA_3}{2A_3}\gamma^a\zeta
-\frac{f_a}{240}\gamma^a\sigma^{(1,0,0)}\zeta
-\frac{g_a}{96}\gamma^a\sigma^{(0,1,1)}\ast\zeta
+\frac{ih_a}{32}\gamma^a\ast\zeta=0,\\\nonumber
\CD_a\zeta+\frac{f_b}{240}\gamma^b\gamma_a\sigma^{(1,0,0)}\zeta-
\frac{g_b}{96}\gamma_a{}^b\sigma^{(0,1,1)}\ast\zeta+
\frac{g_a}{32}\sigma^{(0,1,1)}\ast\zeta-
\frac{ih_b}{96}\gamma_a{}^b\ast\zeta+
\frac{ih_a}{32}\ast\zeta=0
\eeqa

\subsection{Killing vectors, Lorentz rotations and constraints on Spinor bilinears}

The supersymmetry algebra implies that the spinor bilinears 
\beq
\xi^M=\Re(\bar\epsilon_1\gamma^M\epsilon_2)
\eeq
are Killing vectors which generate the Bosonic symmetries that appear 
in the anticommutator (\ref{anticommutator})
of the two supersymmetries generated by $\epsilon_1$ and $\epsilon_2$ \cite{Halmagyi:2005pn}.
This implies the equations
\beq\label{bilinearcond1}\begin{array}{rcl}
-2\Im(\bar\epsilon_1\gamma^\mu\epsilon_2)&=&
\frac{iA_1}{2}\hat\epsilon_1^{\alpha\dot\alpha}\hat\epsilon_2^{\beta\dot\beta}
\left(\bar\chi^{(1,1,1)}_{\alpha\dot\alpha}
((\gamma^{(1)}\gamma^\mu)\otimes\Bid\otimes\Bid)
\chi^{(1,1,1)}_{\beta\dot\beta}\right),\\
-2\Im(\bar\epsilon_1\gamma^m\epsilon_2)&=&
\frac{A_2}{2}\hat\epsilon_1^{\alpha\dot\alpha}\hat\epsilon_2^{\beta\dot\beta}
\left(\bar\chi^{(1,1,1)}_{\alpha\dot\alpha}
(\gamma^{(1)}\otimes\gamma^m\otimes\Bid)
\chi^{(1,1,1)}_{\beta\dot\beta}\right),\\
-2\Im(\bar\epsilon_1\gamma^i\epsilon_2)&=&
\frac{A_3}{2}\hat\epsilon_1^{\alpha\dot\alpha}\hat\epsilon_2^{\beta\dot\beta}
\left(\bar\chi^{(1,1,1)}_{\alpha\dot\alpha}
(\gamma^{(1)}\otimes\Bid\otimes\gamma^i)
\chi^{(1,1,1)}_{\beta\dot\beta}\right),\\
-2\Im(\bar\epsilon_1\gamma^a\epsilon_2)&=&0.
\end{array}\eeq
Furthermore the Lorentz rotation that appears in the anticommutator 
of the two supersymmetries generated by $\epsilon_1$ and $\epsilon_2$ 
is
\beq\label{bilinearcond2}\begin{array}{rl}
l^{MN}=&\omega_P{}^{MN}\xi^P-
\frac{1}{3}F^{MNPQR}\Re(\bar\epsilon_1\gamma_{PQR}\epsilon_2)-\\
&\frac{3}{4}\Im\left(G^{MNP}\bar\epsilon_1\gamma_P\ast\epsilon_2-
\frac{1}{18}G_{PQR}\bar\epsilon_1\gamma^{MNPQR}\ast\epsilon_2\right).
\end{array}\eeq
Comparison with (\ref{anticommutator}) leads to
\beq\begin{array}{rcl}
l^{\mu\nu}&=&\frac{i}{4}\hat\epsilon_1^{\alpha\dot\alpha}\hat\epsilon_2^{\beta\dot\beta}
\left(\bar\chi^{(1,1,1)}_{\alpha\dot\alpha}
((\gamma^{(1)}\gamma^{\mu\nu})\otimes\Bid\otimes\Bid)
\chi^{(1,1,1)}_{\beta\dot\beta}\right),\\
l^{mn}&=&-\frac{i}{4}\hat\epsilon_1^{\alpha\dot\alpha}\hat\epsilon_2^{\beta\dot\beta}
\left(\bar\chi^{(1,1,1)}_{\alpha\dot\alpha}
((\gamma^{(1)}\otimes\gamma^{mn})\otimes\Bid)
\chi^{(1,1,1)}_{\beta\dot\beta}\right),\\
l^{ij}&=&-\frac{i}{4}\hat\epsilon_1^{\alpha\dot\alpha}\hat\epsilon_2^{\beta\dot\beta}
\left(\bar\chi^{(1,1,1)}_{\alpha\dot\alpha}
((\gamma^{(1)}\otimes\Bid\otimes\gamma^{ij}))
\chi^{(1,1,1)}_{\beta\dot\beta}\right)
\end{array}\eeq
with all other Lorentz rotations vanishing.

The left hand sides of (\ref{bilinearcond1}) can be expanded using 
(\ref{killingspinor}), the 
identities (\ref{reducebilinears}) and the symmetry properties
(\ref{bilinearsymmetries}). This implies\footnote{The $\sigma$-s 
are Pauli matrices acting on the $\eta$-indices. Here $\sigma^0=\Bid$ and
$\sigma^{(i,j,k)}=\sigma^i\otimes\sigma^j\otimes\sigma^k$.}
\beq\begin{array}{ll}
\bar\zeta\zeta=A_1,\qquad&
\bar\zeta\sigma^{(1,1,0)}\zeta=\bar\zeta\sigma^{(1,0,1)}\zeta=
\bar\zeta\sigma^{(0,1,1)}\zeta=0,\\
\bar\zeta\sigma^{(2,3,0)}\zeta=A_2,\qquad&
\bar\zeta\sigma^{(3,3,0)}\zeta=\bar\zeta\sigma^{(2,2,0)}\zeta=
\bar\zeta\sigma^{(3,2,0)}\zeta=0,\\
\bar\zeta\sigma^{(2,1,3)}\zeta=A_3,\qquad&
\bar\zeta\sigma^{(3,1,3)}\zeta=\bar\zeta\sigma^{(2,1,2)}\zeta=
\bar\zeta\sigma^{(3,1,2)}\zeta=0,\\ \multicolumn{2}{l}{
\bar\zeta\gamma^a\sigma^{(3,1,0)}\zeta=\bar\zeta\gamma^a\sigma^{(3,0,1)}\zeta=
\bar\zeta\gamma^a\sigma^{(2,1,0)}\zeta=\bar\zeta\gamma^a\sigma^{(2,0,1)}\zeta=
0.}
\end{array}\eeq
Using the chirality condition (\ref{chiralitycond}) one can see that in 
the last set of equations one can set $a=8$. This leaves 16 real equations
for the 16 real components of $\zeta$. The overall phase of $\zeta$ cannot
be determined from those equations. This means that the system of equations 
is overdetermined by one equation.
Doing a cyclic permutation of the Pauli matrices $(0,1,2,3)\rightarrow(0,3,1,2)$ 
the above equations are solved by 
\beq\begin{array}{ll}
f_{1,1,1,1}=e^{i\phi}\alpha,\quad&f_{1,1,-1,1}=i\eta_1e^{i\phi}\alpha^\ast,\quad\\
f_{1,-1,1,1}=-\eta_1\eta_2e^{i\phi}\alpha^\ast,\quad&f_{1,-1,-1,1}=i\eta_2e^{i\phi}\alpha,\\
f_{-1,1,1,1}=-i\eta_2e^{i\phi}\beta,\quad&f_{-1,1,-1,1}=-\eta_1\eta_2e^{i\phi}\beta^\ast,\quad\\
f_{-1,-1,1,1}=-i\eta_1e^{i\phi}\beta^\ast,\quad&f_{-1,-1,-1,1}=e^{i\phi}\beta
\end{array}\eeq
with
\beq\label{normcond}
|\alpha|^2+|\beta|^2=\frac{A_1}{16},\qquad \alpha\beta=\frac{\nu_1(A_2-i\nu_2A_3)}{32}.
\eeq
This leaves $\phi$ and the relative phase of $\alpha$ and $\beta$ 
undetermined. From now on we will continue working in the basis with cyclically 
permuted Pauli matrices.

The conditions (\ref{bilinearcond2}) for $(MN)=(\mu m)$ and $(MN)=(\mu i)$ are
\beq
\Im\left(g_a\bar\epsilon_1\gamma^{\mu 4 5 i a}\ast\epsilon_2\right)=
\Re\left(h_a\bar\epsilon_1\gamma^{\mu m 6 7 a}\ast\epsilon_2\right)=0.
\eeq
These imply the reality conditions
\beq
\Re(e^{-2i\phi}g_a)=\Re(e^{-2i\phi}h_a)=0
\eeq
or
\beq
g_a^\ast=-e^{-4i\phi}g_a\qquad{\rm and}\qquad
h_a^\ast=-e^{-4i\phi}h_a.
\eeq
The other conditions from the closure of the supersymmetry algebra are more involved and
we do not need them.

\subsection{The BPS equations}

We can insert the results of the last section into the Gravitino and Dilatino variation
equations. The dilatino variation equations impose a reality condition on $P$
\beq
(p_8^\ast-ip_9^\ast)=e^{-8i\phi}(p_8-ip_9)
\eeq
one can gauge fix the $U(1)$ R-symmetry of type IIB supergravity by demanding $e^{2i\phi}=i$. Then 
the reality conditions read
\beq
g_a^\ast=g_a,\qquad h_a^\ast=h_a\qquad{\rm and}\qquad
(p_8^\ast-ip_9^\ast)=(p_8-ip_9).
\eeq

To derive the remaining BPS equations it is useful to fix reparametrization invariance by
going to the conformally flat metric on $M_2$
\beq
e^8=A_4\,dx\qquad{\rm and}\qquad e^9=A_4\,dy,
\eeq
and introducing the complex coordinate $z$ by 
\beq
dz=dx+idy.
\eeq
It is useful to combine the real 1-forms $f_a$, $g_a$, $h_a$ and $p_a$ into
\beq\begin{array}{ll}
f=\frac{A_4}{2}(f_8-if_9),\qquad & g=\frac{A_4}{2}(f_8-if_9),\\
h=\frac{A_4}{2}(h_8-ih_9),\qquad{\rm and}\qquad & p=\frac{A_4}{2}(p_8-ip_9).
\end{array}\eeq

The dilatino variation equations then give rise to the BPS equations
\beq\label{bpsdv}\begin{array}{l}
p\beta^\ast+\frac{1}{24}(g+ih)\alpha=0,\\
p\alpha-\frac{1}{24}(g-ih)\beta^\ast=0.
\end{array}\eeq
The Gravitino variation equations in the $\mu$, $m$ and $i$ directions give rise to the BPS equations
\beq\label{bpsgv1}\begin{array}{l}
\frac{\nu_2A_4}{2A_1}\beta+\frac{\partial_zA_1}{A_1}\alpha+\frac{f}{120}\alpha+\frac{g+ih}{48}\beta^\ast=0,\\
\frac{\nu_2A_4}{2A_1}\alpha^\ast-\frac{\partial_zA_1}{A_1}\beta^\ast+\frac{f}{120}\beta^\ast+\frac{g-ih}{48}\alpha=0,\\
-\frac{\nu_1\nu_2A_4}{2A_2}\alpha^\ast+\frac{\partial_zA_2}{A_2}\alpha-\frac{f}{120}\alpha-\frac{3g-ih}{48}\beta^\ast=0,\\
-\frac{\nu_1\nu_2A_4}{2A_2}\beta-\frac{\partial_zA_2}{A_2}\beta^\ast-\frac{f}{120}\beta^\ast-\frac{3g+ih}{48}\alpha=0,\\
\frac{i\nu_1A_4}{2A_3}\alpha^\ast+\frac{\partial_zA_3}{A_3}\alpha-\frac{f}{120}\alpha+\frac{g-3ih}{48}\beta^\ast=0,\\
-\frac{i\nu_1A_4}{2A_3}\beta-\frac{\partial_zA_3}{A_3}\beta^\ast-\frac{f}{120}\beta^\ast+\frac{g+3ih}{48}\alpha=0.
\end{array}\eeq
%{\bf(Check the signs of the first terms in the last four equations!)}
Finally, the gravitino variation equations in the $a$-direction give rise to the reality condition
\beq
q_a=\partial_a\phi,
\eeq
which reduces to $q_a=0$ in the chosen gauge, together with the BPS equations
\beq\label{bpsgv2}\begin{array}{l}
\partial_z\alpha+\frac{\partial_zA_4}{2A_4}\alpha-\frac{g+ih}{48}\beta^\ast=0,\\
\partial_z\beta^\ast+\frac{\partial_zA_4}{2A_4}\beta^\ast+\frac{g-ih}{48}\alpha=0,\\
\partial_z\alpha^\ast-\frac{\partial_zA_4}{2A_4}\alpha^\ast+\frac{f}{120}\alpha^\ast-\frac{g-ih}{24}\beta=0,\\
\partial_z\beta-\frac{\partial_zA_4}{2A_4}\beta-\frac{f}{120}\beta+\frac{g+ih}{24}\alpha^\ast=0.
\end{array}\eeq

\subsection{Bianchi Identities}

The Bianchi identities
\beq\begin{array}{c}
\CD P=0,\\
\CD G=-P\wedge G^\ast,\\
dQ=-i P\wedge P^\ast,\\
dF=\frac{5i}{12}G\wedge G^\ast
\end{array}\eeq
turn into the equations
\beq\begin{array}{c}
dp^{(4)}=0,\\
d(A_2^2\,g^{(4)})+p^{(4)}\wedge(A_2^2\,g^{(4)})=0,\\
d(A_3^2\,h^{(4)})-p^{(4)}\wedge(A_3^2\,h^{(4)})=0,\\
p^{(4)}\wedge p^{(4)}{}^\ast=0,\\
d(A_1^4\,f^{(4)})=0,\\
d(A_2^2\,A_3^2\,\ast f^{(4)})=\frac{5}{6}\,A_2^2\,A_3^2\,g^{(4)}\wedge h^{(4)}.
\end{array}\eeq
The first four identities can be solved by introducing the functions
$\rho$, $l$, $m$ and $n$
\beq\label{bianchi2}\begin{array}{c}
p^{(4)}=d\rho,\\
g^{(4)}=\frac{e^{-\rho}}{A_2^2}\,dm,\\
h^{(4)}=\frac{e^{\rho}}{A_3^2}\,dn,\\
f^{(4)}=\frac{1}{A_1^4}\,dl,
\end{array}\eeq
the last equation leads to a harmonic equation for $l$.

\section{The domain wall geometries}\label{secwallgeometries}

In this section we will attempt to 'solve' the system (\ref{bpsdv}), (\ref{bpsgv1}) 
and (\ref{bpsgv2}) of BPS equations. Those equations are real linear in $\alpha$ 
and $\beta$. For this reason we can rescale $\alpha$ and $\beta$ such that the 
normalization conditions (\ref{normcond}) are nicer
\beq\label{normcond2}
A_1=\alpha\alpha^\ast+\beta\beta^\ast,\qquad
A_2=\nu_1(\alpha\beta+\alpha^\ast\beta^\ast)\qquad{\rm and}\qquad
A_3=i\nu_1\nu_2(\alpha\beta-\alpha^\ast\beta^\ast).
\eeq
Those conditions will turn out to be crucial for ``bootstrapping" the system.

In order to get a better understanding of what to expect from the general solution,
let us first start by verifying that $AdS_5\times S^5$ is a solution and where 
supersymmetric brane probes are sitting.

\subsection{$AdS_5\times S^5$}

The pure $AdS_5\times S^5$ solution is an $AdS_4\times S^2\times S^2$ fibration over
an infinite strip. On the one boundary of the strip one $S^2$ is shrinking to zero 
size, whereas on the other boundary the other $S^2$ is shrinking to zero size. 
This can be seen by embedding $AdS_5$ into $\BR^6$ spanned by $X_{-1}, X_0\cdots,X_4$ and
$S^5$ into $\BR^6$ spanned by $Y_1,\cdots,Y_6$
\beq
 -X_{-1}^2-X_0^2+X_1^2+\cdots+X_4^2=-R^2\qquad{\rm and}\qquad
Y_1^2+\cdots+Y_6^2=R^2.
\eeq
Then the strip can be parametrized by $-\infty<X_4<\infty$ and $0<r<R$ such that
\beq
Y_1^2+Y_2^2+Y_3^2=r^2\qquad{\rm and}\qquad Y_4^2+Y_5^2+Y_6^2=R^2-r^2.
\eeq

The solution has no 3-form flux, i.e. $g=h=0$ and 
the dilatino variation equations (\ref{bpsdv}) imply that the dilaton is
constant $p=0$. The gravitino variation equations (\ref{bpsgv2}) lead to
the holomorphicity conditions
\beq
\partial_{\bar z}(\alpha^\ast{}^2A_4)=\partial_{\bar z}(\beta{}^2A_4)=
\partial_{\bar z}\frac{\alpha\beta^\ast}{A_4}=0.
\eeq
This implies that $|\alpha|^2|\beta|^2$ is holomorphic and real, i.e.
\beq
|\alpha|^2|\beta|^2=c^4.
\eeq
Furthermore, $\alpha\beta|\beta|^2$ is holomorphic and (\ref{normcond2}) implies that it 
is real on one boundary of $M_2$ and imaginary on the other one. This 
determines\footnote{One could choose a different holomorphic function, but the 
solution would still locally be $AdS_5\times S^5$.}
\beq
\alpha\beta|\beta|^2=c^4e^{z},
\eeq
which implies
\beq
\alpha=ce^{-\frac{x}{2}+i\phi_\alpha}\qquad{\rm and}\qquad
\beta=ce^{\frac{x}{2}+iy-i\phi_\alpha}.
\eeq
Using the equations (\ref{normcond2}) we can determine
\beq
A_1=2c^2\cosh(x),\qquad
A_2=2c^2\nu_1\cos(y)\qquad{\rm and}\qquad
A_3=-2c^2\nu_1\nu_2\sin(y).
\eeq
For the range of $y\in[0,\frac{\pi}{2}]$ to make sense, we have to set $\nu_1=1$
and $\nu_2=-1$. The gravitino variation equations (\ref{bpsgv1}) then imply
\beq
\alpha=ce^{-\frac{z^\ast}{2}},\qquad\beta=ce^{\frac{z}{2}},\qquad 
A_4=2c^2\qquad{\rm and}\qquad f=60.
\eeq
From this we conclude that 
\beq
2c^2=R.
\eeq

The general solutions that we are looking for have to have this asymptotic form, 
i.e. they have to have semi-infinite strips where $A_2=0$ on one side and $A_3=0$ on 
the other side with a constant dilaton and no 3-form fluxes.

%\subsection{'Pure' D5-branes?}

%Pure D5-branes don't change the rank of the gauge group. A sufficient condition
%for this is $G\wedge G^\ast=0$ or
%\beq
%h=t g,\qquad{\rm where}\qquad t\in\BR.
%\eeq
%Then the dilatino variation equations (\ref{bpsdv}) imply
%\beq
%|\alpha|=|\beta|.
%\eeq
%This is in contradiction with the asymptotic $AdS_5\times S^5$ behaviour. Imposing 
%the D5-brane projector is even stronger and leads to much worse inconsistencies

\subsection{The general 'bootstrap'}\label{secbootstr}

%The signs $\nu_1$ and $\nu_2$ are just conventions. In order to conform with
%the $AdS_5\times S^5$ solution that we presented, we set $\nu_1=1$ and 
%$\nu_2=-1$.
Let us start by using (\ref{bpsdv}) 
\beq
\frac{g}{12}=p\left(\frac{\alpha}{\beta^\ast}-\frac{\beta^\ast}{\alpha}\right),\qquad
\frac{ih}{12}=-p\left(\frac{\alpha}{\beta^\ast}+\frac{\beta^\ast}{\alpha}\right)
\eeq 
and (\ref{normcond2}) to eliminate $g$, $h$, $A_1$, $A_2$ and $A_3$ 
from the BPS equations. The equations (\ref{bpsgv1}) allow to solve for
$A_4$, $f$, $p$ in terms of $\alpha$ and $\beta$
\beq\begin{array}{l}
\frac{\nu_2A_4}{2\alpha\beta^\ast}=
\frac{4\frac{\alpha\beta}{\alpha^\ast\beta^\ast}-4\frac{\alpha^\ast\beta^\ast}{\alpha\beta}}
{\left(\frac{\alpha}{\beta^\ast}\right)^2+\left(\frac{\beta^\ast}{\alpha}\right)^2-
2\frac{\alpha\beta}{\alpha^\ast\beta^\ast}-2\frac{\alpha^\ast\beta^\ast}{\alpha\beta}}
\,\partial_z\log|\alpha\beta|-
\partial_z\log\left(\frac{\alpha\beta}{\alpha^\ast\beta^\ast}\right),\\
p=-\frac{4}{\left(\frac{\alpha}{\beta^\ast}\right)^2+\left(\frac{\beta^\ast}{\alpha}\right)^2-
2\frac{\alpha\beta}{\alpha^\ast\beta^\ast}-2\frac{\alpha^\ast\beta^\ast}{\alpha\beta}}\,
\partial_z\log|\alpha\beta|,\\
\frac{f}{60}=\partial_z\log\left(\frac{\alpha\beta}{\alpha^\ast\beta^\ast}\right)+
2\frac{\left(\frac{\alpha}{\beta^\ast}\right)^2-\left(\frac{\beta^\ast}{\alpha}\right)^2-
2\frac{\alpha\beta}{\alpha^\ast\beta^\ast}+2\frac{\alpha^\ast\beta^\ast}{\alpha\beta}}
{\left(\frac{\alpha}{\beta^\ast}\right)^2+\left(\frac{\beta^\ast}{\alpha}\right)^2-
2\frac{\alpha\beta}{\alpha^\ast\beta^\ast}-2\frac{\alpha^\ast\beta^\ast}{\alpha\beta}}
\,\partial_z\log|\alpha\beta|
\end{array}\eeq
and lead to one more independent equation for $\alpha$ and $\beta$
\beq\label{bootstr1}
\frac{|\alpha|^2-|\beta|^2}{|\alpha|^2+|\beta|^2}\,\frac{\nu_2A_4}{2\alpha\beta^\ast}-
2\partial_z\log(|\alpha|^2+|\beta|^2)+
\frac{p}{2}\left(\left(\frac{\alpha}{\beta^\ast}\right)^2+\left(\frac{\beta^\ast}{\alpha}\right)^2\right)=0.
\eeq

The difference of the first two equations (\ref{bpsgv2}) leads to the identity
\beq\label{bootstr2}
\frac{\left(\frac{\alpha}{\beta^\ast}\right)^2+\left(\frac{\beta^\ast}{\alpha}\right)^2-
2\frac{\alpha\beta}{\alpha^\ast\beta^\ast}-2\frac{\alpha^\ast\beta^\ast}{\alpha\beta}}
{\left(\frac{\alpha}{\beta^\ast}\right)^2-\left(\frac{\beta^\ast}{\alpha}\right)^2}\,
\partial_z\log\left(\frac{\alpha}{\beta^\ast}\right)=-2\partial_z\log|\alpha\beta|
\eeq
and the sum of the first two equations (\ref{bpsgv2}) leads to
\beq
\partial_z\left(\frac{\alpha^2\beta^\ast{}^2A_4^2}
{\left(\frac{\alpha}{\beta^\ast}\right)^2-\left(\frac{\beta^\ast}{\alpha}\right)^2}\right)=0,
\eeq
which implies that 
\beq\label{bootstr3}
\frac{\alpha^2\beta^\ast{}^2A_4^2}
{\left(\frac{\alpha}{\beta^\ast}\right)^2-\left(\frac{\beta^\ast}{\alpha}\right)^2}=a(z^\ast),
\eeq
where $a(z^\ast)$ is an antiholomorphic function. The other two equations (\ref{bpsgv2}) are redundant.

We can reexpress $\frac{\nu_2A_4}{2\alpha\beta^\ast}$ and $p$ in terms of the ratio
$\frac{\alpha}{\beta^\ast}$
\beq\label{a4p}\begin{array}{l}
\frac{\nu_2A_4}{2\alpha\beta^\ast}=
-2\frac{\frac{\alpha\beta}{\alpha^\ast\beta^\ast}-\frac{\alpha^\ast\beta^\ast}{\alpha\beta}}
{\left(\frac{\alpha}{\beta^\ast}\right)^2-\left(\frac{\beta^\ast}{\alpha}\right)^2}\,
\partial_z\log\left(\frac{\alpha}{\beta^\ast}\right)-
\partial_z\log\left(\frac{\alpha\beta}{\beta^\ast\alpha^\ast}\right),\\
p=\frac{2}{\left(\frac{\alpha}{\beta^\ast}\right)^2-\left(\frac{\beta^\ast}{\alpha}\right)^2}\,
\partial_z\log\left(\frac{\alpha}{\beta^\ast}\right).
\end{array}\eeq
The equations (\ref{bootstr1}), (\ref{bootstr2}) and (\ref{bootstr3}) are then the remaining system
of equations for $\alpha$ and $\beta$. Actually, (\ref{bootstr1}) only depends on the ratio
$\frac{\alpha}{\beta^\ast}$ and turns out to be the last equation that is trivially satisfied.
The other two equations form a second order system for $\alpha$ and $\beta$.

Those equations are algebraic in the phase of $\alpha\beta^\ast$. Equation (\ref{bootstr3})
can be solved for $|\alpha\beta^\ast|$ in terms of $\frac{\alpha}{\beta^\ast}$, this can be inserted 
into (\ref{bootstr2}) to give a single second order differential equation for 
$\frac{\alpha}{\beta^\ast}$.

%Now, the second equation (\ref{a4p}) together with the Bianchi identity for $p$ can be used to
%express $\frac{\alpha}{\beta^\ast}$ as
%\beq
%\left(\frac{\alpha}{\beta^\ast}\right)^2=\tanh(\rho-a(z^\ast)),
%\eeq
%where $a(z^\ast)$ is an antiholomorphic function of $z$, which can be chosen to be $a(z^\ast)=z^\ast$.

\subsection{Probe branes and boundary conditions}\label{secprobebc}

To start understanding the general solution let us first look at the probe branes again.
The projector equation for a supersymmetric NS5-brane around $S^2$ with $k$ units 
of magnetic flux is \cite{Gauntlett:1997cv}
\beq
\frac{i}{\sqrt{1+\left(\frac{\pi k}{R}\right)^2}}\,\gamma^{(1)}\gamma^{(2)}\ast\epsilon-
\frac{\frac{\pi k}{R}}{\sqrt{1+\left(\frac{\pi k}{R}\right)^2}}\,\gamma^{(1)}\epsilon=\epsilon,
\eeq
and the projector equation for a supersymmetric D5-brane around $\tilde S^2$ with $k$ units 
of magnetic flux is
\beq
-\frac{1}{\sqrt{1+\left(\frac{\pi k}{R}\right)^2}}\,\gamma^{(1)}\gamma^{(3)}\ast\epsilon-
\frac{\frac{\pi k}{R}}{\sqrt{1+\left(\frac{\pi k}{R}\right)^2}}\,\gamma^{(1)}\epsilon=\epsilon.
\eeq
Those equations turn into
\beq
e^{z_{NS5_k}^\ast}=\frac{\beta^\ast}{\alpha}=\sqrt{1+\left(\frac{\pi k}{R}\right)^2}+\frac{\pi k}{R}
\quad{\rm and}\quad
e^{z_{D5_k}^\ast}=\frac{\beta^\ast}{\alpha}=-i\left(\sqrt{1+\left(\frac{\pi k}{R}\right)^2}+\frac{\pi k}{R}\right)
\eeq
which is
\beq
\sinh(x(k))=\frac{\pi k}{R},\qquad y_{NS5}=0\qquad{\rm or}\qquad y_{D5}=\frac{\pi}{2}
\eeq
in agreement with the predictions of section \ref{secprobe}.
From this it is easy to see that the NS5-branes are sitting in a place where 
$\frac{\alpha}{\beta^\ast}$ is real, whereas the D5-branes are sitting in a place where
$\frac{\alpha}{\beta^\ast}$ is imaginary. The absolute value $\left|\frac{\alpha}{\beta^\ast}\right|$
determines the magnetic flux on the brane.

For the $AdS_5\times S^5$ solution this means that the NS5-branes are sitting on the boundary
of the strip where $S^2$ has maximal size, call it the 'black' boundary and the 
D5-branes are sitting on the boundary of the strip where $\tilde S^2$ has maximal size, 
call it the 'white' boundary.
\begin{figure}[bth]
\centerline{ \epsfig{file=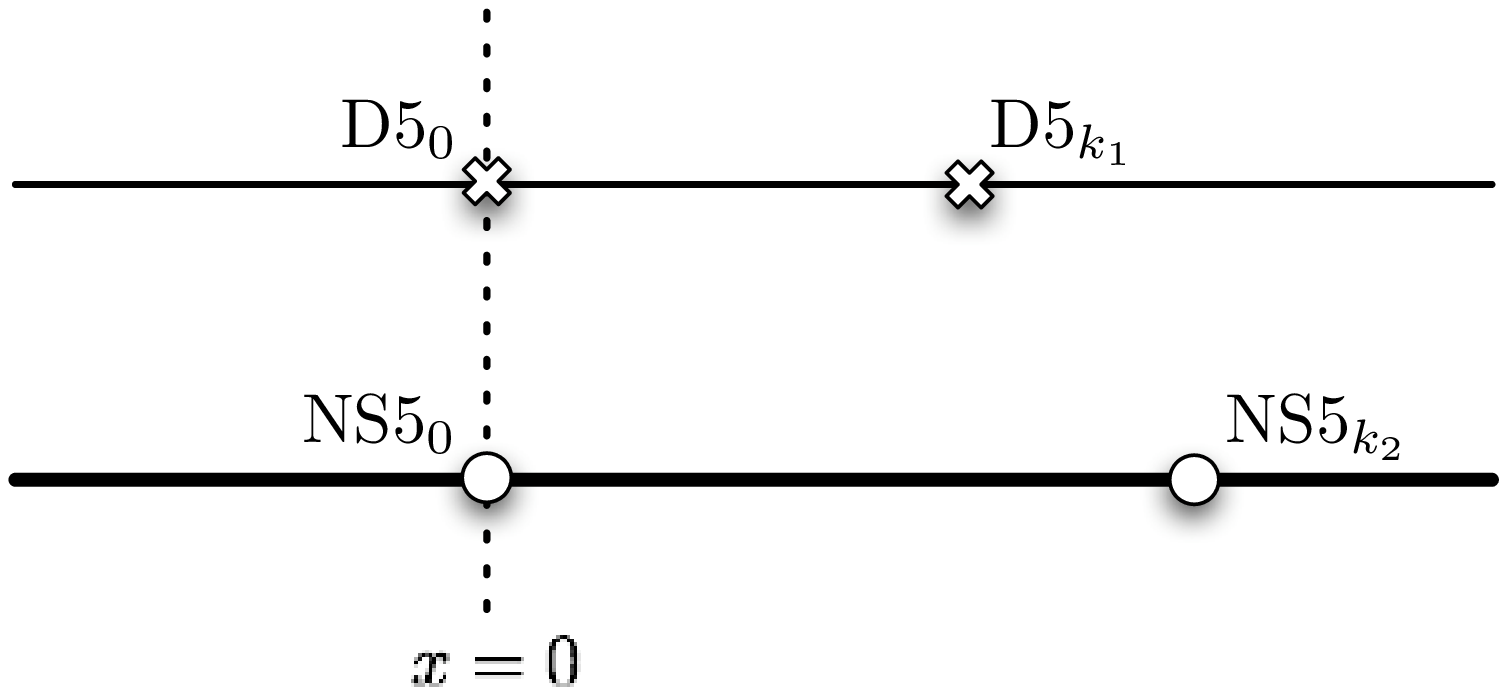,width=12cm}}
\caption{\sl Probe branes in $AdS_5\times S^5$.}
\end{figure}
In the regions where the 5-branes are sitting, we expect a backreaction of the geometry, 
which generates the throat of a 5-brane. This means that there is a 3-sphere which supports 
the appropriate 3-form flux. This is done by switching the shrunk 2-sphere in the respective
region of the boundary. To understand this better, we need to work out the boundary 
conditions in the different regions.

The two dimensional geometry can be conformally mapped to a region in the complex plane. This region 
has a boundary on which one of the two 2-spheres is shrinking to zero size. In order to parametrize
the boundary in a more invariant way, we impose the boundary condition
\beq
A_4|_{\partial M_2}=1.
\eeq
The boundary is divided into (colored) segments on which either one or the other 2-sphere is 
shrinking to zero size
\beq
A_2|_{\partial M_2,w}=0\qquad{\rm or}\qquad A_3|_{\partial M_2,b}=0.
\eeq
Using (\ref{normcond2}) this leads to the same conditions on the phase of 
$\frac{\alpha}{\beta^\ast}$ as the five-brane projectors do.
Furthermore, in order for the geometry to be smooth, one has to require either
\beq
\partial_nA_1|_{\partial M_2,w}=0,\qquad\partial_nA_2|_{\partial M_2,w}=1,\qquad
\partial_nA_3|_{\partial M_2,w}=0,\qquad{\rm and}\qquad\partial_nA_4|_{\partial M_2,w}=0,
\eeq
or alternatively
\beq
\partial_nA_1|_{\partial M_2,b}=0,\qquad\partial_nA_2|_{\partial M_2,b}=0,\qquad
\partial_nA_3|_{\partial M_2,b}=1,\qquad{\rm and}\qquad\partial_nA_4|_{\partial M_2,b}=0,
\eeq
where $\partial_n$ is the normal derivative to the boundary. In order for the fluxes to
be regular, we either need to require
\beq
p_n|_{\partial M_2,w}=f_n|_{\partial M_2,w}=g_t|_{\partial M_2,w}=h_n|_{\partial M_2,w}=0
\eeq
or
\beq
p_n|_{\partial M_2,b}=f_n|_{\partial M_2,b}=g_n|_{\partial M_2,b}=h_t|_{\partial M_2,b}=0.
\eeq
It is not difficult to see that the boundary conditions on $g$ and $h$ are satisfied, once
the boundary condition on $p$ is satisfied.

Let us concentrate on the 'white' boundary, where $A_2|_{\partial M_2,w}=0$. We assume that
it is along the $x$-axis and that the strip is on the upper half plane. There the boundary 
conditions imply for the spinor variables $\frac{\alpha}{\beta^\ast}$ and $\alpha\beta^\ast$
\beq\begin{array}{lcl}
\frac{\alpha}{\beta^\ast}|_{\partial M_2,w}\in i\BR,\qquad
\alpha\beta^\ast|_{\partial M_2,w}\in \BR,&\qquad&
\left(\frac{\alpha}{\beta^\ast}\,\left|\frac{\beta^\ast}{\alpha}\right|\,
\frac{|\alpha\beta^\ast|}{\alpha\beta^\ast}\right)_{\partial M_2,w}=i\nu_1\nu_2,\\
\partial_y\left(\frac{\alpha}{\beta^\ast}\right)_{\partial M_2,w}=
\frac{\nu_1}{2}
\left(\left|\frac{\alpha}{\beta^\ast}\right|\frac{1}{|\alpha\beta^\ast|}\right)_{\partial M_2,w},&&
\partial_y|\alpha\beta^\ast|_{\partial M_2,w}=0.
\end{array}\eeq
Note that all the normal derivatives of the spinor variables are determined, except for 
$\partial_y\arg(\alpha\beta^\ast)|_{\partial M_2,w}$. This phase only appears in the expression 
for $A_4$ in terms of the spinor variables.

The antiholomorphic function $a(z^\ast)$ has to be real on this boundary. Given $a(z^\ast)$ and
using both boundary conditions on $A_4$, (\ref{bootstr3}) can be solved for $|\alpha\beta^\ast|$
\beq
|\alpha\beta^\ast|_{\partial M_2,w}=
|a|\left|\left(\frac{\alpha}{\beta^\ast}\right)^2-\left(\frac{\beta^\ast}{\alpha}\right)^2\right|.
\eeq
This can be inserted into (\ref{bootstr2}) to give a second order ODE for $\frac{\alpha}{\beta^\ast}$.
The solutions of that ODE are determined by the values of $\frac{\alpha}{\beta^\ast}$ at
the 'ends' of the 'white' boundary. The above boundary conditions are then enough for the second 
order PDE of \ref{secbootstr}.
The antiholomorphic function presumably has to be determined by the reality condition above and
its asymptotic behavior.

Similarly the boundary conditions on the spinor variables for a 'black' boundary along the y-axis, 
where the strip is the right half plane are
\beq\begin{array}{lcl}
\frac{\alpha}{\beta^\ast}|_{\partial M_2,b}\in \BR,\qquad
\alpha\beta^\ast|_{\partial M_2,b}\in i\BR,&\qquad&
\left(\frac{\alpha}{\beta^\ast}\,\left|\frac{\beta^\ast}{\alpha}\right|\,
\frac{|\alpha\beta^\ast|}{\alpha\beta^\ast}\right)_{\partial M_2,b}=i\nu_1,\\
\partial_x\left(\frac{\alpha}{\beta^\ast}\right)_{\partial M_2,b}=
-\frac{i\nu_1\nu_2}{2}
\left(\left|\frac{\alpha}{\beta^\ast}\right|\frac{1}{|\alpha\beta^\ast|}\right)_{\partial M_2,b},&&
\partial_x|\alpha\beta^\ast|_{\partial M_2,b}=0.
\end{array}\eeq

Let us try to see how this story fits in with the probe brane picture. We expect that the five-branes
get replaced by geometry with fluxes. At the position of the defect we expect that the value of
$\frac{\alpha}{\beta^\ast}$ agrees with the one from the probe brane calculation. There are
two possibilities that can happen: The defect is either a finite or an infinite distance along 
the boundary away from a given reference point. Furthermore the geometry at the defect 
has to have a 3-cycle that supports the flux.

For a defect at finite distance this can be done by a change of coloring, i.e. by inserting 
a finite interval of 'white' boundary into the 'black' boundary or vice versa. This creates 
a 3-sphere which can support the flux. On the 'black' side of the interface, the value of 
$\frac{\alpha}{\beta^\ast}$ is given by the probe brane value.
\begin{figure}[bth]
\centerline{ \epsfig{file=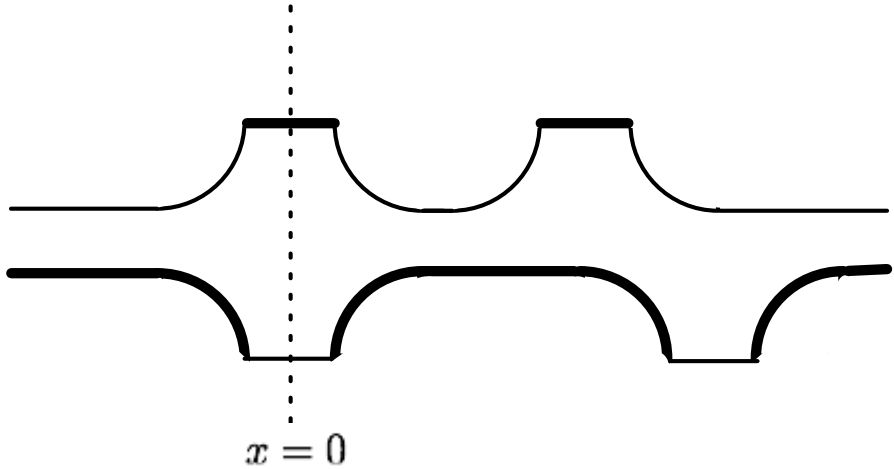,width=12cm}}
\caption{\sl A possibility for the backreacted geometry.}
\end{figure}

In order for the geometry to be smooth at the interface of a 'black' and a 'white' boundary, 
it needs to have a right angle, such that the strip turns
locally into a quadrant. Unlike the cases of chiral operators or Wilson lines 
\cite{Lin:2004nb,Yamaguchi:2006te,Lunin:2006xr}, there are no such interfaces in our vacuum 
($AdS_5\times S^5$) solution. Actually, closer examination reveals that not all the 
regularity conditions can hold at the same time. 
For example, if the boundary conditions on $A_2$ and $A_3$ hold at the same time, then 
$\partial_z\log|\alpha\beta|$ diverges at the interface. This implies that the dilaton $p$
diverges or $\frac{\alpha}{\beta^\ast}$ diverges.

On the other hand a defect at infinite distance produces an infinite throat with the same
color on both sides. This also creates a three-sphere to support the three-form flux. We expect that 
$\frac{\alpha}{\beta^\ast}$ asymptotes to the value given by the probe brane picture.

In the asymptotic region of such a throat we expect $\frac{\alpha}{\beta^\ast}$ to be
almost constant at the boundary. This implies that $\alpha\beta^\ast$ is linearly growing 
at the boundary, i.e. the warp factors $A_1$ and $A_3$ are growing linearly and the dilaton 
is growing logarithmically along the boundary. 
In the other direction this is of course bounded and cannot continue forever, it has to
connect to an asymptotically $AdS_5\times S^5$ region.

As of now we haven't found any convincing argument for either scenario, but those seem 
to be the only possibilities for the backreacted geometry of a five-brane.

We believe that those difficulties are arising due to the fact that we are trying to describe 
the backreacted geometry of five-branes instead of D3-branes \cite{Lin:2004nb} 
or strings \cite{Yamaguchi:2006te,Lunin:2006xr}, where the geometry 
seems to be really well behaved at the locations of the 'defects'.

The gravity discussion also 
leaves open the possiblity of more than two asymptotic $AdS_5\times S^5$ regions. This would 
correspond to several $\CN=4$ super Yang Mills theories that interact on a defect. In the case 
of only a single asymptotic $AdS_5\times S^5$ region one would get a $\CN=4$ super Yang Mills 
theory with a boundary. The latter case requires an interface between a 'black' and a 'white' 
boundary.

There is more work to be done in order to understand those outstanding issues better. To complete 
the story, one also needs to calculate the fluxes through all the three-cycles as well as change of 
the rank of the gauge group. Those impose the true physical boundary conditions and might be 
calculable even if the full solution is not known (see e.g. \cite{Halmagyi:2005pn}). We leave a 
closer examination of all those issues for future work \cite{wip}.

\begin{acknowledgments} 
\nopagebreak

\noindent 
We would like to thank Sujay Ashok, Eleonora Dell'Aquila, Jerome Gauntlett, Roberto Emparan, Anton Kapustin and Rob Myers for useful discussions. Research at the Perimeter Institute is supported in part by funds from NSERC of Canada and by MEDT of Ontario. JG is further supported by an NSERC Discovery grant.

\end{acknowledgments}

\begin{appendix}

\section{Clifford algebra conventions}\label{appclifford}

\subsection{Generalities}

The Clifford algebra is defined by the anticommutation relations
\beq
\{\gamma^m,\gamma^n\}=2\eta^{mn},
\eeq
where $\eta^{mn}=\eta^m\delta^{mn}$. We choose a representation in which 
$\sqrt{\eta^m}\gamma^m$ is Hermitean\footnote{By the square root we mean 
$\sqrt{1}=1$ and $\sqrt{-1}=i$.}. Given a complex structure, one can 
define the raising and lowering operators
\beq
\Gamma^m=\sqrt{\eta^{2m}}\gamma^{2m}+i\sqrt{\eta^{2m+1}}\gamma^{2m+1},
\quad{\rm and}\quad (\Gamma^m)^\dagger=
\sqrt{\eta^{2m}}\gamma^{2m}-i\sqrt{\eta^{2m+1}}\gamma^{2m+1}.
\eeq
Then the raising and lowering operators satisfy the following anticommutation 
relations:
\beq
\{\Gamma^m,\Gamma^n\}=\{(\Gamma^m)^\dagger,(\Gamma^n)^\dagger\}=0
\quad{\rm and}\quad \{\Gamma^m,(\Gamma^n)^\dagger\}=4\delta^{mn}.
\eeq
One can then define the fermion number operators
\beq
F^m=i\sqrt{\eta^{2m}}\gamma^{2m}\sqrt{\eta^{2m+1}}\gamma^{2m+1}=
1-\half\Gamma^m(\Gamma^m)^\dagger=-1+\half(\Gamma^m)^\dagger\Gamma^m.
\eeq
The chirality operator is then the product of all the Fermion number operators
$\gamma=F^1\cdots F^n$.

The Fermion number operators have eigenvalues $\pm 1$. The eigenvalues of
the Fermion number operators can be used to label a basis of states.
One can define a ground state $\ket{0}$ which is anihilated by all the 
lowering operators. It has Fermion number $-1$ for all Fermion number 
operators. All other states can be gotten by applying raising operators.
If one labels a state by $\ket{\nu_1,\cdots,\nu_n}$, then the raising and
lowering operators act as follows:
\beqa
\ket{\nu_1,\cdots,+1,\cdots,\nu_n}&=&
\half\nu_1\cdots\nu_{m-1}(\Gamma^m)^\dagger
\ket{\nu_1,\cdots,-1,\cdots,\nu_n},\\
\ket{\nu_1,\cdots,-1,\cdots,\nu_n}&=&
\half\nu_1\cdots\nu_{m-1}\Gamma^m
\ket{\nu_1,\cdots,+1,\cdots,\nu_n}.
\eeqa
This defines the matrix elements of the gamma matrices. One can see that in 
this basis $\Gamma^m$ is real. From this follows that
\begin{itemize}
\item the matrices $\sqrt{\eta^m}\gamma^m$ are Hermitean,
\item the matrices $\sqrt{\eta^{2m}}\gamma^{2m}$ are symmetric and real and
\item the matrices $\sqrt{\eta^{2m+1}}\gamma^{2m+1}$ are antisymmetric and 
imaginary.
\end{itemize}

In general there are matrices $B$, $C$ and $D$ such that
\beqa
(\gamma^m)^\ast&=&\eta_BB\gamma^m B^{-1},\\
(\gamma^m)^\dagger&=&\eta_CC\gamma^m C^{-1},\\
(\gamma^m)^t&=&\eta_DD\gamma^m D^{-1},
\eeqa
where $\eta_B,\eta_C,\eta_D=\pm 1$ is a constant. Given a spinor $\epsilon$, 
$\ast\epsilon=B^{-1}\epsilon^\ast$, $\bar\epsilon=\epsilon^\dagger C$ and 
$\tilde\epsilon=\epsilon^t D$ transform covariantly.

If $BB^\ast=\Bid$ one can impose the Majorana condition 
$\epsilon=\ast\epsilon$.
And if $B$ commutes with the chirality operator $\gamma$, one can impose 
the Majorana-Weyl condition.

\subsection{$Spin(1,9)$}

Chirality operator:
\beq
\gamma^{(10)}=\gamma^{0\cdots 9}
\eeq

Complex conjugation:
\beq
B^{(10)}=\gamma^{013579}
\eeq

\beq
B^{(10)}\gamma^M(B^{(10)})^{-1}=(\gamma^M)^\ast
\eeq

Hermitean conjugation:
\beq
C^{(10)}=\gamma^0
\eeq

\beq
C^{(10)}\gamma^M(C^{(10)})^{-1}=-(\gamma^M)^\dagger
\eeq

Transpose:
\beq
D^{(10)}=\gamma^{13579}
\eeq

\beq
D^{(10)}\gamma^M(D^{(10)})^{-1}=-(\gamma^M)^t
\eeq

\subsection{$Spin(1,3)$ -- $AdS_4$}

\beq
\gamma^{(1)}=i\gamma^{0123}
\eeq

Complex Conjugation:
\beq
B^{(1)}=\gamma^{2}
\eeq

\beq
B^{(1)}\gamma^\mu (B^{(1)})^{-1}=(\gamma^\mu)^\ast\qquad{\rm and}\qquad 
B^{(1)}(i\gamma^{(1)})(B^{(1)})^{-1}=(i\gamma^{(1)})^\ast
\eeq

\beq
B^{(1)}(B^{(1)})^\ast=\Bid
\eeq

Hermitean conjugation:
\beq
C^{(1)}=\gamma^{123}
\eeq

\beq
C^{(1)}\gamma^\mu (C^{(1)})^{-1}=(\gamma^\mu)^\dagger\qquad{\rm and}\qquad 
C^{(1)}(i\gamma^{(1)})(C^{(1)})^{-1}=(i\gamma^{(1)})^\dagger
\eeq

Transpose:
\beq
D^{(1)}=\gamma^{13}
\eeq

\beq
D^{(1)}\gamma^\mu (D^{(1)})^{-1}=(\gamma^\mu)^t\qquad{\rm and}\qquad 
D^{(1)}(i\gamma^{(1)})(D^{(1)})^{-1}=(i\gamma^{(1)})^t
\eeq

\subsection{$Spin(2)$ -- $S^2$}

\beq
\gamma^{(2)}=i\gamma^{45}
\eeq

Complex Conjugation:
\beq
B^{(2)}=\gamma^{5}
\eeq

\beq
B^{(2)}\gamma^m (B^{(2)})^{-1}=-(\gamma^m)^\ast\qquad{\rm and}\qquad 
B^{(2)}\gamma^{(2)}(B^{(2)})^{-1}=-(\gamma^{(2)})^\ast
\eeq

\beq
B^{(2)}(B^{(2)})^*=-\Bid
\eeq

Hermitean conjugation:
\beq
C^{(2)}=\Bid
\eeq

\beq
C^{(2)}\gamma^m(C^{(2)})^{-1}=(\gamma^m)^\dagger\qquad{\rm and}\qquad 
C^{(2)}\gamma^{(2)}(C^{(2)})^{-1}=(\gamma^{(2)})^\dagger
\eeq

Transpose:
\beq
D^{(2)}=\gamma^{5}
\eeq

\beq
D^{(2)}\gamma^m(D^{(2)})^{-1}=-(\gamma^m)^t\qquad{\rm and}\qquad 
D^{(2)}\gamma^{(2)}(D^{(2)})^{-1}=-(\gamma^{(2)})^t
\eeq

Similarly 
\beq
\gamma^{(3)}=i\gamma^{67},\qquad 
B^{(3)}=\gamma^7,\qquad
C^{(3)}=\Bid\qquad{\rm and}\qquad
D^{(3)}=\gamma^7.
\eeq

\subsection{$Spin(2)$ -- $M_2$}

\beq
\gamma^{(2)}=i\gamma^{89}
\eeq

Complex Conjugation:
\beq
B^{(4)}=\gamma^{8}
\eeq

\beq
B^{(4)}\gamma^a (B^{(4)})^{-1}=(\gamma^a)^\ast
\eeq

\beq
B^{(4)}(B^{(4)})^*=\Bid
\eeq

Hermitean conjugation:
\beq
C^{(4)}=\Bid
\eeq

\beq
C^{(4)}\gamma^a(C^{(4)})^{-1}=(\gamma^m)^\dagger
\eeq

Transpose:
\beq
D^{(4)}=\gamma^{8}
\eeq

\beq
D^{(4)}\gamma^a(D^{(4)})^{-1}=(\gamma^a)^t
\eeq

\subsection{Decomposition of a 10-dimensional Spinor}

The ten dimensional gamma matrix algebra can be decomposed in the folowing way
\beq\begin{array}{l}
\gamma^\mu=\gamma^\mu\otimes\Bid\otimes\Bid\otimes\Bid,\\
\gamma^m=\gamma^{(1)}\otimes\gamma^m\otimes\Bid\otimes\Bid,\\
\gamma^i=\gamma^{(1)}\otimes\gamma^{(2)}\otimes\gamma^i\otimes\Bid,\\
\gamma^a=\gamma^{(1)}\otimes\gamma^{(2)}\otimes\gamma^{(3)}\otimes\gamma^a,\\
\gamma^{(10)}=
\gamma^{(1)}\otimes\gamma^{(2)}\otimes\gamma^{(3)}\otimes\gamma^{(4)},\\
B^{(10)}=-B^{(1)}\otimes B^{(2)}\otimes(B^{(3)}\gamma^{(3)})
\otimes(B^{(4)}\gamma^{(4)}),\\
C^{(10)}=
-i(C^{(1)}\gamma^{(1)})\otimes C^{(2)}\otimes C^{(3)}\otimes C^{(4)},\\
D^{(10)}=-i(D^{(1)}\gamma^{(1)})\otimes D^{(2)}\otimes(D^{(3)}\gamma^{(3)})
\otimes(D^{(4)}\gamma^{(4)}).
\end{array}\eeq

\section{Relations for spinor bilinears}

In this appendix we summarize some properties of spinor bilinears.
\beqa
(\bar\epsilon_2\gamma^M\epsilon_1)^\dagger&=&
\bar\epsilon_1\gamma^M\epsilon_2,\\
\tilde\epsilon_1\gamma^M\epsilon_2&=&
-\tilde\epsilon_2\gamma^M\epsilon_1,\\
\overline{\ast\epsilon_1}\gamma^M\ast\epsilon_2&=&
\bar\epsilon_2\gamma^M\epsilon_1,\\
\overline{\ast\epsilon_1}\gamma^M\epsilon_2&=&
\tilde\epsilon_1\gamma^M\epsilon_2.
\eeqa
The following table summarizes the symmetry properties of 8-dimensional spinor
bilinears under transpositio(exchange of $(\alpha\dot\alpha)$ and $(\beta\dot\beta)$)
and complex conjugation
\beq\label{bilinearsymmetries}\begin{array}{|l|l|l|}
\hline
{\rm bilinear}&t&\ast\\\hline
\bar\chi^{(1,1,1)}_{\alpha\dot\alpha}
((\gamma^{(1)}{}^{\frac{\eta_1+\eta_1^\prime}{2}}\gamma^\mu)
\otimes(\gamma^{(2)}{}^{\frac{\eta_2-\eta_2^\prime}{2}})
\otimes(\gamma^{(3)}{}^{\frac{\eta_3-\eta_3^\prime}{2}}))
\chi^{(1,1,1)}_{\beta\dot\beta}&
\eta_1\eta_1^\prime\eta_2\eta_2^\prime\eta_3\eta_3^\prime& -\eta\eta^\prime\\
\bar\chi^{(1,1,1)}_{\alpha\dot\alpha}
((\gamma^{(1)}{}^{\frac{\eta_1-\eta_1^\prime}{2}})
\otimes(\gamma^{(2)}{}^{\frac{\eta_2-\eta_2^\prime}{2}}\gamma^m)
\otimes(\gamma^{(3)}{}^{\frac{\eta_3-\eta_3^\prime}{2}}))
\chi^{(1,1,1)}_{\beta\dot\beta}&
\eta_3\eta_3^\prime& -\eta\eta^\prime\\
\bar\chi^{(1,1,1)}_{\alpha\dot\alpha}
((\gamma^{(1)}{}^{\frac{\eta_1-\eta_1^\prime}{2}})
\otimes(\gamma^{(2)}{}^{\frac{\eta_2+\eta_2^\prime}{2}})
\otimes(\gamma^{(3)}{}^{\frac{\eta_3-\eta_3^\prime}{2}}\gamma^i))
\chi^{(1,1,1)}_{\beta\dot\beta}&
-\eta_2\eta_2^\prime& \eta\eta^\prime\\
\bar\chi^{(1,1,1)}_{\alpha\dot\alpha}
((\gamma^{(1)}{}^{\frac{\eta_1-\eta_1^\prime}{2}})
\otimes(\gamma^{(2)}{}^{\frac{\eta_2+\eta_2^\prime}{2}})
\otimes(\gamma^{(3)}{}^{\frac{\eta_3+\eta_3^\prime}{2}}))
\chi^{(1,1,1)}_{\beta\dot\beta}&
-\eta_2\eta_2^\prime\eta_3\eta_3^\prime& \eta\eta^\prime\\
\bar\chi^{(1,1,1)}_{\alpha\dot\alpha}
((\gamma^{(1)}{}^{\frac{\eta_1+\eta_1^\prime}{2}}\gamma^{\mu\nu\rho})
\otimes(\gamma^{(2)}{}^{\frac{\eta_2-\eta_2^\prime}{2}})
\otimes(\gamma^{(3)}{}^{\frac{\eta_3-\eta_3^\prime}{2}}))
\chi^{(1,1,1)}_{\beta\dot\beta}&
-\eta_1\eta_1^\prime\eta_2\eta_2^\prime\eta_3\eta_3^\prime& -\eta\eta^\prime\\
\bar\chi^{(1,1,1)}_{\alpha\dot\alpha}
((\gamma^{(1)}{}^{\frac{\eta_1-\eta_1^\prime}{2}}\gamma^{\mu\nu})
\otimes(\gamma^{(2)}{}^{\frac{\eta_2+\eta_2^\prime}{2}})
\otimes(\gamma^{(3)}{}^{\frac{\eta_3+\eta_3^\prime}{2}}))
\chi^{(1,1,1)}_{\beta\dot\beta}&
\eta_2\eta_2^\prime\eta_3\eta_3^\prime& \eta\eta^\prime\\
\bar\chi^{(1,1,1)}_{\alpha\dot\alpha}
((\gamma^{(1)}{}^{\frac{\eta_1-\eta_1^\prime}{2}})
\otimes(\gamma^{(2)}{}^{\frac{\eta_2+\eta_2^\prime}{2}}\gamma^{mn})
\otimes(\gamma^{(3)}{}^{\frac{\eta_3+\eta_3^\prime}{2}}))
\chi^{(1,1,1)}_{\beta\dot\beta}&
\eta_2\eta_2^\prime\eta_3\eta_3^\prime& \eta\eta^\prime\\
\bar\chi^{(1,1,1)}_{\alpha\dot\alpha}
((\gamma^{(1)}{}^{\frac{\eta_1-\eta_1^\prime}{2}})
\otimes(\gamma^{(2)}{}^{\frac{\eta_2+\eta_2^\prime}{2}})
\otimes(\gamma^{(3)}{}^{\frac{\eta_3+\eta_3^\prime}{2}}\gamma^{ij}))
\chi^{(1,1,1)}_{\beta\dot\beta}&
\eta_2\eta_2^\prime\eta_3\eta_3^\prime& \eta\eta^\prime\\
\bar\chi^{(1,1,1)}_{\alpha\dot\alpha}
((\gamma^{(1)}{}^{\frac{\eta_1-\eta_1^\prime}{2}})
\otimes(\gamma^{(2)}{}^{\frac{\eta_2-\eta_2^\prime}{2}}\gamma^m)
\otimes(\gamma^{(3)}{}^{\frac{\eta_3+\eta_3^\prime}{2}}\gamma^i))
\chi^{(1,1,1)}_{\beta\dot\beta}&
-1& -\eta\eta^\prime\\
\bar\chi^{(1,1,1)}_{\alpha\dot\alpha}
((\gamma^{(1)}{}^{\frac{\eta_1-\eta_1^\prime}{2}})
\otimes(\gamma^{(2)}{}^{\frac{\eta_2+\eta_2^\prime}{2}}\gamma^{mn})
\otimes(\gamma^{(3)}{}^{\frac{\eta_3-\eta_3^\prime}{2}}\gamma^i))
\chi^{(1,1,1)}_{\beta\dot\beta}&
\eta_2\eta_2^\prime& \eta\eta^\prime\\
\bar\chi^{(1,1,1)}_{\alpha\dot\alpha}
((\gamma^{(1)}{}^{\frac{\eta_1-\eta_1^\prime}{2}})
\otimes(\gamma^{(2)}{}^{\frac{\eta_2-\eta_2^\prime}{2}}\gamma^m)
\otimes(\gamma^{(3)}{}^{\frac{\eta_3-\eta_3^\prime}{2}}\gamma^{ij}))
\chi^{(1,1,1)}_{\beta\dot\beta}&
-\eta_3\eta_3^\prime& -\eta\eta^\prime\\
\bar\chi^{(1,1,1)}_{\alpha\dot\alpha}
((\gamma^{(1)}{}^{\frac{\eta_1-\eta_1^\prime}{2}}\gamma^{\mu\nu})
\otimes(\gamma^{(2)}{}^{\frac{\eta_2+\eta_2^\prime}{2}}\gamma^{mn})
\otimes(\gamma^{(3)}{}^{\frac{\eta_3-\eta_3^\prime}{2}}\gamma^i))
\chi^{(1,1,1)}_{\beta\dot\beta}&
-\eta_2\eta_2^\prime& \eta\eta^\prime\\
\bar\chi^{(1,1,1)}_{\alpha\dot\alpha}
((\gamma^{(1)}{}^{\frac{\eta_1-\eta_1^\prime}{2}}\gamma^{\mu\nu})
\otimes(\gamma^{(2)}{}^{\frac{\eta_2-\eta_2^\prime}{2}}\gamma^m)
\otimes(\gamma^{(3)}{}^{\frac{\eta_3-\eta_3^\prime}{2}}\gamma^{ij}))
\chi^{(1,1,1)}_{\beta\dot\beta}&
\eta_3\eta_3^\prime& -\eta\eta^\prime\\
\bar\chi^{(1,1,1)}_{\alpha\dot\alpha}
((\gamma^{(1)}{}^{\frac{\eta_1+\eta_1^\prime}{2}}\gamma^\mu)
\otimes(\gamma^{(2)}{}^{\frac{\eta_2-\eta_2^\prime}{2}}\gamma^{mn})
\otimes(\gamma^{(3)}{}^{\frac{\eta_3-\eta_3^\prime}{2}}\gamma^{ij}))
\chi^{(1,1,1)}_{\beta\dot\beta}&
\eta_1\eta_1^\prime\eta_2\eta_2^\prime\eta_3\eta_3^\prime& -\eta\eta^\prime\\
\bar\chi^{(1,1,1)}_{\alpha\dot\alpha}
((\gamma^{(1)}{}^{\frac{\eta_1-\eta_1^\prime}{2}}\gamma^{\mu\nu})
\otimes(\gamma^{(2)}{}^{\frac{\eta_2+\eta_2^\prime}{2}}\gamma^{mn})
\otimes(\gamma^{(3)}{}^{\frac{\eta_3+\eta_3^\prime}{2}}))
\chi^{(1,1,1)}_{\beta\dot\beta}&
-\eta_2\eta_2^\prime\eta_3\eta_3^\prime& \eta\eta^\prime\\
\bar\chi^{(1,1,1)}_{\alpha\dot\alpha}
((\gamma^{(1)}{}^{\frac{\eta_1-\eta_1^\prime}{2}}\gamma^{\mu\nu})
\otimes(\gamma^{(2)}{}^{\frac{\eta_2+\eta_2^\prime}{2}})
\otimes(\gamma^{(3)}{}^{\frac{\eta_3+\eta_3^\prime}{2}}\gamma^{ij}))
\chi^{(1,1,1)}_{\beta\dot\beta}&
-\eta_2\eta_2^\prime\eta_3\eta_3^\prime& \eta\eta^\prime\\
\bar\chi^{(1,1,1)}_{\alpha\dot\alpha}
((\gamma^{(1)}{}^{\frac{\eta_1+\eta_1^\prime}{2}}\gamma^\mu)
\otimes(\gamma^{(2)}{}^{\frac{\eta_2-\eta_2^\prime}{2}}\gamma^{mn})
\otimes(\gamma^{(3)}{}^{\frac{\eta_3+\eta_3^\prime}{2}}\gamma^i))
\chi^{(1,1,1)}_{\beta\dot\beta}&
\eta_1\eta_1^\prime\eta_2\eta_2^\prime& -\eta\eta^\prime\\
\bar\chi^{(1,1,1)}_{\alpha\dot\alpha}
((\gamma^{(1)}{}^{\frac{\eta_1+\eta_1^\prime}{2}}\gamma^\mu)
\otimes(\gamma^{(2)}{}^{\frac{\eta_2+\eta_2^\prime}{2}}\gamma^m)
\otimes(\gamma^{(3)}{}^{\frac{\eta_3+\eta_3^\prime}{2}}\gamma^{ij}))
\chi^{(1,1,1)}_{\beta\dot\beta}&
-\eta_1\eta_1^\prime\eta_3\eta_3^\prime& \eta\eta^\prime\\
\bar\chi^{(1,1,1)}_{\alpha\dot\alpha}
((\gamma^{(1)}{}^{\frac{\eta_1+\eta_1^\prime}{2}}\gamma^\mu)
\otimes(\gamma^{(2)}{}^{\frac{\eta_2-\eta_2^\prime}{2}}\gamma^{mn})
\otimes(\gamma^{(3)}{}^{\frac{\eta_3-\eta_3^\prime}{2}}))
\chi^{(1,1,1)}_{\beta\dot\beta}&
-\eta_1\eta_1^\prime\eta_2\eta_2^\prime\eta_3\eta_3^\prime& -\eta\eta^\prime\\
\bar\chi^{(1,1,1)}_{\alpha\dot\alpha}
((\gamma^{(1)}{}^{\frac{\eta_1+\eta_1^\prime}{2}}\gamma^\mu)
\otimes(\gamma^{(2)}{}^{\frac{\eta_2-\eta_2^\prime}{2}})
\otimes(\gamma^{(3)}{}^{\frac{\eta_3-\eta_3^\prime}{2}}\gamma^{ij}))
\chi^{(1,1,1)}_{\beta\dot\beta}&
-\eta_1\eta_1^\prime\eta_2\eta_2^\prime\eta_3\eta_3^\prime& -\eta\eta^\prime\\
\bar\chi^{(1,1,1)}_{\alpha\dot\alpha}
((\gamma^{(1)}{}^{\frac{\eta_1-\eta_1^\prime}{2}})
\otimes(\gamma^{(2)}{}^{\frac{\eta_2+\eta_2^\prime}{2}}\gamma^{mn})
\otimes(\gamma^{(3)}{}^{\frac{\eta_3-\eta_3^\prime}{2}}\gamma^i))
\chi^{(1,1,1)}_{\beta\dot\beta}&
\eta_2\eta_2^\prime& \eta\eta^\prime\\
\bar\chi^{(1,1,1)}_{\alpha\dot\alpha}
((\gamma^{(1)}{}^{\frac{\eta_1-\eta_1^\prime}{2}})
\otimes(\gamma^{(2)}{}^{\frac{\eta_2-\eta_2^\prime}{2}}\gamma^m)
\otimes(\gamma^{(3)}{}^{\frac{\eta_3-\eta_3^\prime}{2}}\gamma^{ij}))
\chi^{(1,1,1)}_{\beta\dot\beta}&
-\eta_3\eta_3^\prime& -\eta\eta^\prime\\
\hline
\end{array}\eeq
where $\eta\eta^\prime=\eta_1\eta_1^\prime\eta_2\eta_2^\prime\eta_3\eta_3^\prime$.

\end{appendix}

\vfill
\pagebreak
 
\bibliographystyle{utphys}
\bibliography{wall}

\end{document}